\documentclass[journal]{IEEEtran}

\usepackage{cite}
\ifCLASSINFOpdf
\usepackage[pdftex]{graphicx}
\else
\usepackage[dvips]{graphicx}
\DeclareGraphicsExtensions{.eps}
\fi
\graphicspath{{fig/pdf/}}
\usepackage{import}

\usepackage{amsmath}
\usepackage{mathtools}

\usepackage{siunitx}

\usepackage[ruled, noline, noend, linesnumbered]{algorithm2e}
\SetCommentSty{font=regular}
\newcommand{\pluseq}{\mathrel{+}=}
\let\oldnl\nl
\newcommand{\nonl}{\renewcommand{\nl}{\let\nl\oldnl}}

\usepackage{array}
\usepackage[table]{xcolor}
\usepackage{booktabs}
\usepackage{multirow}
\usepackage{threeparttable}

\newcommand{\STAB}[1]{\begin{tabular}{@{}c@{}}#1\end{tabular}}
\definecolor{lg}{gray}{0.85}
\usepackage{comment}

\ifCLASSOPTIONcaptionsoff
\usepackage[nomarkers]{endfloat}
\let\MYoriglatexcaption\caption
\renewcommand{\caption}[2][\relax]{\MYoriglatexcaption[#2]{#2}}
\fi
\usepackage[spaces,hyphens]{xurl} 
\usepackage[colorlinks,allcolors=blue]{hyperref}

\usepackage{subfiles}
\usepackage{xr}
\usepackage{tikz}
\newcommand*\circled[1]{\tikz[baseline=(char.base)]{
            \node[shape=circle,draw,inner sep=.7pt] (char) {#1};}}
\newcommand*\smallcircled[1]{\tikz[baseline=(char.base)]{
            \node[shape=circle,draw,inner sep=.5pt] (char) {#1};}}

\newcommand{\etal}{\textit{et al.}}

\begin{document}
%
\title{Kraken: An Efficient Engine with a \\Uniform Dataflow for Deep Neural Networks}

\author{G.~Abarajithan and Chamira~U.~S.~Edussooriya,~\IEEEmembership{Member,~IEEE}
\thanks{G~Abarajithan is with the Department of Electronic and Telecommunication Engineering, University of Moratuwa, Moratuwa 10400, Sri Lanka, (e-mail: abarajithang@uom.lk).}
\thanks{C. U. S. Edussooriya is with the Department of Electronic and Telecommunication Engineering, University of Moratuwa, Moratuwa 10400, Sri Lanka and the Department of Electrical and Computer Engineering, Florida International University, Miami, FL, USA (e-mail: chamira@uom.lk).}}

\maketitle

\newcommand{\vConvFreqMhz}{400}
\newcommand{\vFcFreqMhz  }{200}

\newcommand{\vR}{7}
\newcommand{\vC}{96}
\newcommand{\vNumPEs}{672}
\newcommand{\vTechNm}{65}
\newcommand{\vArea}{7.3}
\newcommand{\vGates}{1111k}
\newcommand{\vRam}{384.0}
\newcommand{\vGopsPeak}{537.6}
\newcommand{\vGopsPerArea}{73.32}
\newcommand{\vGopsPerW}{512.16}
\newcommand{\vAreaPEMacPercent}{87.12}


\newcommand{\vNumPETimesEyeriss    }{4}
\newcommand{\vMemoryTimesEyeriss   }{2.1}
\newcommand{\vAreaTimesEyeriss     }{0.6}
\newcommand{\vNumPETimesZascad     }{3.5}
\newcommand{\vMemoryTimesZascad    }{10.4}
\newcommand{\vAreaTimesZascad      }{1.2}
\newcommand{\vNumPETimesCarla      }{3.4}
\newcommand{\vMemoryTimesCarla     }{4.5}
\newcommand{\vAreaTimesCarla       }{1.2}
\newcommand{\vFreqTimesCarla       }{2}
\newcommand{\vGopsPerAreaTimesCarla}{5.8}
\newcommand{\vGopsPerWTimesCarla   }{1.6}


\newcommand{\vAlexMacsZeros}{669.7 M}
\newcommand{\vAlexMacsValid}{616.2 M}
\newcommand{\vAlexMacsFc   }{55.5 M}
\newcommand{\vVggMacsZeros}{15.3 G}
\newcommand{\vVggMacsValid}{14.8 G}
\newcommand{\vVggMacsFc   }{123.6 M}
\newcommand{\vResMacsZeros}{3.9 G}
\newcommand{\vResMacsValid}{3.7 G}
\newcommand{\vResMacsFc   }{2.0 M}




 \newcommand{\vAlexConvPEf    }{77.2}
 \newcommand{\vAlexConvFps    }{336.6}
 \newcommand{\vAlexConvLatency}{3.0}
 \newcommand{\vAlexConvBatch  }{1}
 \newcommand{\vAlexConvMAMn   }{6.4}
 \newcommand{\vAlexConvMAMb   }{7.5}
 \newcommand{\vAlexConvGops   }{414.8}
 \newcommand{\vAlexConvGopsAr }{56.6}
 \newcommand{\vAlexConvAI     }{191.8}
 \newcommand{\vAlexConvPower  }{1050}
 \newcommand{\vAlexConvEnEf   }{395.2}

 \newcommand{\vAlexFcPEf    }{99.1}
 \newcommand{\vAlexFcFps    }{2.4k}
 \newcommand{\vAlexFcLatency}{2.9}
 \newcommand{\vAlexFcBatch  }{7}
 \newcommand{\vAlexFcMAMn   }{12.2}
 \newcommand{\vAlexFcMAMb   }{11.7}
 \newcommand{\vAlexFcGops   }{266.5}
 \newcommand{\vAlexFcGopsAr }{36.3}
 \newcommand{\vAlexFcAI     }{9.1}
 \newcommand{\vAlexFcPower  }{613}
 \newcommand{\vAlexFcEnEf   }{434.8}


 \newcommand{\vVggConvPEf    }{96.5}
 \newcommand{\vVggConvFps    }{17.5}
 \newcommand{\vVggConvLatency}{57.2}
 \newcommand{\vVggConvBatch  }{1}
 \newcommand{\vVggConvMAMn   }{96.8}
 \newcommand{\vVggConvMAMb   }{118.1}
 \newcommand{\vVggConvGops   }{518.7}
 \newcommand{\vVggConvGopsAr }{70.7}
 \newcommand{\vVggConvAI     }{306.8}
 \newcommand{\vVggConvPower  }{1050}
 \newcommand{\vVggConvEnEf   }{494.1}

 \newcommand{\vVggFcPEf    }{99.1}
 \newcommand{\vVggFcFps    }{1.1k}
 \newcommand{\vVggFcLatency}{6.5}
 \newcommand{\vVggFcBatch  }{7}
 \newcommand{\vVggFcMAMn   }{27.0}
 \newcommand{\vVggFcMAMb   }{25.9}
 \newcommand{\vVggFcGops   }{266.3}
 \newcommand{\vVggFcGopsAr }{36.3}
 \newcommand{\vVggFcAI     }{9.2}
 \newcommand{\vVggFcPower  }{613}
 \newcommand{\vVggFcEnEf   }{434.5}


 \newcommand{\vResConvPEf    }{88.3}
 \newcommand{\vResConvFps    }{64.2}
 \newcommand{\vResConvLatency}{15.6}
 \newcommand{\vResConvBatch  }{1}
 \newcommand{\vResConvMAMn   }{67.9}
 \newcommand{\vResConvMAMb   }{87.5}
 \newcommand{\vResConvGops   }{474.9}
 \newcommand{\vResConvGopsAr }{64.8}
 \newcommand{\vResConvAI     }{108.9}
 \newcommand{\vResConvPower  }{1050}
 \newcommand{\vResConvEnEf   }{452.4}

 \newcommand{\vResFcPEf    }{94.7}
 \newcommand{\vResFcFps    }{62.1k}
 \newcommand{\vResFcLatency}{0.1}
 \newcommand{\vResFcBatch  }{7}
 \newcommand{\vResFcMAMn   }{0.5}
 \newcommand{\vResFcMAMb   }{0.5}
 \newcommand{\vResFcGops   }{254.5}
 \newcommand{\vResFcGopsAr }{34.7}
 \newcommand{\vResFcAI     }{8.6}
 \newcommand{\vResFcPower  }{613}
 \newcommand{\vResFcEnEf   }{415.3}



 \newcommand{\vAlexZascadConvPEf    }{66.4}
 \newcommand{\vAlexZascadConvFps    }{48.1}
 \newcommand{\vAlexZascadConvLatency}{20.8}
 \newcommand{\vAlexZascadConvBatch  }{1}
 \newcommand{\vAlexZascadConvMAMn   }{8.7}
 \newcommand{\vAlexZascadConvMAMb   }{16.6}
 \newcommand{\vAlexZascadConvGops   }{59.3}
 \newcommand{\vAlexZascadConvGopsAr }{9.9}
 \newcommand{\vAlexZascadConvAI     }{142.2}
 \newcommand{\vAlexZascadConvPower  }{265}
 \newcommand{\vAlexZascadConvEnEf   }{223.7}

 \newcommand{\vAlexZascadFcPEf    }{96.8}
 \newcommand{\vAlexZascadFcFps    }{131.6}
 \newcommand{\vAlexZascadFcLatency}{7.6}
 \newcommand{\vAlexZascadFcBatch  }{1}
 \newcommand{\vAlexZascadFcMAMn   }{55.8}
 \newcommand{\vAlexZascadFcMAMb   }{117.8}
 \newcommand{\vAlexZascadFcGops   }{14.6}
 \newcommand{\vAlexZascadFcGopsAr }{2.4}
 \newcommand{\vAlexZascadFcAI     }{2.0}
 \newcommand{\vAlexZascadFcPower  }{37}
 \newcommand{\vAlexZascadFcEnEf   }{395.0}


 \newcommand{\vVggZascadConvPEf    }{78.7}
 \newcommand{\vVggZascadConvFps    }{2.2}
 \newcommand{\vVggZascadConvLatency}{421.8}
 \newcommand{\vVggZascadConvBatch  }{1}
 \newcommand{\vVggZascadConvMAMn   }{205.2}
 \newcommand{\vVggZascadConvMAMb   }{375.5}
 \newcommand{\vVggZascadConvGops   }{65.3}
 \newcommand{\vVggZascadConvGopsAr }{10.9}
 \newcommand{\vVggZascadConvAI     }{144.7}
 \newcommand{\vVggZascadConvPower  }{301}
 \newcommand{\vVggZascadConvEnEf   }{217.0}

 \newcommand{\vVggZascadFcPEf    }{96.6}
 \newcommand{\vVggZascadFcFps    }{61.0}
 \newcommand{\vVggZascadFcLatency}{16.4}
 \newcommand{\vVggZascadFcBatch  }{1}
 \newcommand{\vVggZascadFcMAMn   }{124.3}
 \newcommand{\vVggZascadFcMAMb   }{247.3}
 \newcommand{\vVggZascadFcGops   }{15.1}
 \newcommand{\vVggZascadFcGopsAr }{2.5}
 \newcommand{\vVggZascadFcAI     }{2.0}
 \newcommand{\vVggZascadFcPower  }{40}
 \newcommand{\vVggZascadFcEnEf   }{377.1}


 \newcommand{\vResZascadConvPEf    }{51.9}
 \newcommand{\vResZascadConvFps    }{9.6}
 \newcommand{\vResZascadConvLatency}{103.6}
 \newcommand{\vResZascadConvBatch  }{1}
 \newcommand{\vResZascadConvMAMn   }{102.1}
 \newcommand{\vResZascadConvMAMb   }{154.6}
 \newcommand{\vResZascadConvGops   }{71.0}
 \newcommand{\vResZascadConvGopsAr }{11.8}
 \newcommand{\vResZascadConvAI     }{72.4}
 \newcommand{\vResZascadConvPower  }{248}
 \newcommand{\vResZascadConvEnEf   }{286.2}

 \newcommand{\vResZascadFcPEf    }{86.8}
 \newcommand{\vResZascadFcFps    }{3.3k}
 \newcommand{\vResZascadFcLatency}{0.3}
 \newcommand{\vResZascadFcBatch  }{1}
 \newcommand{\vResZascadFcMAMn   }{2.1}
 \newcommand{\vResZascadFcMAMb   }{4.1}
 \newcommand{\vResZascadFcGops   }{13.5}
 \newcommand{\vResZascadFcGopsAr }{2.3}
 \newcommand{\vResZascadFcAI     }{2.0}
 \newcommand{\vResZascadFcPower  }{36}
 \newcommand{\vResZascadFcEnEf   }{380.8}



 \newcommand{\vVggCarlaConvPEf    }{96.4}
 \newcommand{\vVggCarlaConvFps    }{2.5}
 \newcommand{\vVggCarlaConvLatency}{396.9}
 \newcommand{\vVggCarlaConvBatch  }{1}
 \newcommand{\vVggCarlaConvMAMn   }{129.4}
 \newcommand{\vVggCarlaConvMAMb   }{258.2}
 \newcommand{\vVggCarlaConvGops   }{74.2}
 \newcommand{\vVggCarlaConvGopsAr }{12.0}
 \newcommand{\vVggCarlaConvAI     }{229.4}
 \newcommand{\vVggCarlaConvPower  }{247}
 \newcommand{\vVggCarlaConvEnEf   }{300.5}


 \newcommand{\vResCarlaConvPEf    }{89.5}
 \newcommand{\vResCarlaConvFps    }{10.8}
 \newcommand{\vResCarlaConvLatency}{92.7}
 \newcommand{\vResCarlaConvBatch  }{1}
 \newcommand{\vResCarlaConvMAMn   }{69.1}
 \newcommand{\vResCarlaConvMAMb   }{124.0}
 \newcommand{\vResCarlaConvGops   }{79.8}
 \newcommand{\vResCarlaConvGopsAr }{12.9}
 \newcommand{\vResCarlaConvAI     }{107.0}
 \newcommand{\vResCarlaConvPower  }{247}
 \newcommand{\vResCarlaConvEnEf   }{323.3}

\newcommand{\vResFirstPEf    }{73.1}
\newcommand{\vResFirstAltPEf }{79.8}


\begin{abstract}
Deep neural networks (DNNs) have been successfully employed in a multitude of applications with remarkable performance. As such performance is achieved at a significant computational cost, several embedded applications demand fast and efficient hardware accelerators for DNNs. Previously proposed  application specific integrated circuit (ASIC) architectures strive to utilize arrays of hundreds of processing elements (PEs) and reduce power-hungry DRAM accesses using multiple dataflows requiring complex PE architectures. These consume significant area and reduce the maximum clock frequency. This paper introduces the Kraken architecture, which optimally processes the convolutional layers, fully-connected layers, and matrix products of any DNN through a hardware-friendly uniform dataflow. This enables maximal data reuse of weights, inputs, and outputs, with a bare-bones PE design and on-the-fly dynamic reconfiguration. Kraken, implemented in \vTechNm{}-nm CMOS technology at \vConvFreqMhz{} MHz, packs \vNumPEs{} PEs in \vArea{} mm$^2$, with a peak performance of \vGopsPeak{} Gops. Kraken processes the convolutional layers of AlexNet, VGG-16, and ResNet-50 at \vAlexConvFps{}, \vVggConvFps{}, and \vResConvFps{} frames/s, respectively, hence outperforming the state-of-the-art ASIC architectures in terms of overall performance efficiency, DRAM accesses, arithmetic intensity, and throughput, with \vGopsPerAreaTimesCarla{}$\times$ more Gops/mm$^2$ and \vGopsPerWTimesCarla{}$\times$ more Gops/W.

\end{abstract}

\begin{IEEEkeywords}
  Convolutional neural networks (CNNs), 
  deep learning, 
  dataflow processing, 
  energy-efficient accelerators, 
  spatial architecture,
  application specific integrated circuits (ASIC),
  reconfigurable architecture.
\end{IEEEkeywords}

%
\IEEEpeerreviewmaketitle

\section{Introduction}

\IEEEPARstart{D}{eep} Neural Networks (DNNs) have been widely adopted in modern automation systems that require accurate classification and detection due to their remarkable performance in complex pattern recognition tasks. DNNs have been growing deeper and deeper in the past few years, empowering them with beyond human-level capabilities \cite{cnn_survey,imagenet_analysis}. However, this has been achieved at the expense of increased computational complexity, while many mobile and edge processing applications require fast inference of such DNNs with low power and low chip area.

In order to address this rising demand for efficient inference, application-specific  integrated  circuit (ASIC) architectures are developed as arrays of hundreds of processing elements (PEs), where each PE performs a \emph{multiply-accumulate} (MAC) operation. Since convolutional layers, fully connected layers, and matrix products do not impose an order of performing the MACs, the design space for such hardware architectures and their corresponding spatio-temporal orchestrations of data (called dataflows) is quite large. 

The amount of data required for the computation of each layer in modern DNNs is in the order of several megabytes \cite{cnn_survey}, which cannot fit the on-chip memories. While this makes repeated DRAM accesses inevitable, they are primarily responsible for the energy consumption in hardware accelerators. For instance, a 32-bit DRAM access consumes 200 times the energy required for a MAC operation in 45 nm technology \cite{eie}. In addition, a PE array can only perform a few hundred operations in parallel, a tiny subset of the hundreds of millions of operations required for a layer in a DNN. Naively mapping the operations of layers of varying shapes to the fixed hardware architecture would result in PEs idling without work. Therefore, it is a challenging task of paramount importance to design a generic dataflow that maximizes the overall performance efficiency by optimally utilizing the fixed PE array architecture, while reducing the number of memory accesses by exploiting data reuse opportunities, over varying shapes and types of DNN layers.

Several architectures and their corresponding dataflows have been introduced in the literature over the past few years to accelerate the inference of DNNs \cite{survey, dnn_book, updated_survey}. Weight-stationary dataflows such as NVDLA \cite{nvdla}, TPU \cite{tpu}, neuFlow \cite{neuflow}, Sankaradas \etal{} \cite{sankaradas}, Park \etal{} \cite{park}, Chakradhar \etal{} \cite{chakradhar}, Sriram \etal{} \cite{sriram}, Cambricorn-X \cite{cambricon_x} and Origami \cite{origami} hold the weights in register files or SRAMs inside PEs over multiple operations. Input-stationary dataflows such as SCNN \cite{scnn} hold input pixels inside PEs while changing weights. Output-stationary dataflows such as DaDianNao \cite{dadiannao}, DianNao \cite{diannao}, Zhang \etal{} \cite{zhang}, Moons \etal{} \cite{moons}, ShiDianNao \cite{shidiannao}, and Gupta \etal{} \cite{gupta} are designed to minimize the energy consumption of reading and writing partial sums. The DianNao family of accelerators \cite{diannao_family,diannao,dadiannao,shidiannao} minimize memory accesses by storing the entire neural network within their eDRAM buffers and are hence evaluated only on older, smaller networks. 

DNA \cite{dna} supports three different dataflows to individually optimize the reuse of inputs, outputs, and weights, resulting in a complex 3-level PE structure with large multiplexers. Chen \etal{} introduced a row-stationary dataflow with an architecture named Eyeriss \cite{eyeriss_arch,eyeriss}, which maximizes the reuse of weights, inputs and partial sums using scratchpads inside its 168 PEs. The scratchpads take 46.8\% of the total area and 47.9\% of the total power, in addition to the global buffers that take 18.8\% of area. The 2-D array of PEs is controlled through a network-on-chip structure that orchestrates the proposed dataflow to process both convolutional and fully-connected layers. It was benchmarked on AlexNet and VGG-16 among state-of-the-art convolutional neural networks (CNNs) with high arithmetic intensity. However, its dataflow does not efficiently utilize the PE array due to long reconfiguration times and inability to perform computations when transferring data, resulting in low performance efficiency. Jo \etal{} \cite{dsip} introduced DSIP with a master-slave ISA to overlap data transfer and computation. DSIP also employs scratchpads that take 45.9\% of total area and 35.6\% of total power. Ardakani \etal{} introduced the Fully-Connected Inspired Dataflow (FID) \cite{fid} for VGG-like CNNs, where $3{\times}3$ and $1{\times}1$ convolutional layers are treated as special cases of fully-connected layers. FID was then generalized into GFID with corresponding architectures named Multi-Mode Inference Engine (MMIE) \cite{multi_mode} and ZASCAD \cite{zasca} with the ability to process larger filter sizes needed for AlexNet and ResNet-50. While MMIE/ZASCAD reports a high utilization factor (percentage of PEs active in a computational clock cycle), its overall performance efficiency is relatively low due to clock cycles wasted during weight passing and data transfer. Ahmadi \etal{} introduced an architecture \cite{carla_initial} for the convolutional layers of VGG-like CNNs and then generalized it into CARLA \cite{carla}, implemented as an array of 196 PEs. CARLA can process only the convolutional layers of CNNs, and it is tailored for $3{\times}3$ and $1{\times}1$ convolutional layers where the number of output channels is a multiple of 64. This results in a low performance efficiency for convolutional layers with larger filter sizes. CARLA also uses four different dataflows to maximize its utilization, requiring large multiplexers, resulting in mostly idle datapaths in its architecture.

Another parallel area of research is exploiting sparsity in compressed DNNs. The first published versions of Eyeriss \cite{eyeriss}, MMIE \cite{multi_mode}, and Ahmadi \etal{} \cite{carla_initial} focused on dense DNNs. They were later extended to exploit sparsity as Eyeriss v2 \cite{eyeriss_v2}, ZASCAS \cite{zasca} and CARLA \cite{carla}, respectively. Eyeriss v2 and EIE \cite{eie} use the Compressed Sparse Column (CSC) scheme, while Cnvlutin \cite{cnvlutin} uses Compressed Sparse Row (CSR) to exploit the zeros resulting from the Rectified Linear Unit (ReLU) activation function, which is commonly found in early CNNs. ZASCAS \cite{zasca} and EIE are designed to skip such null activations. Cambricorn-X \cite{cambricon_x} exploits sparsity in pruned weights, whereas SCNN \cite{scnn} exploits sparsity in both weights and activations. Moons \etal{} explored the effects of quantization on CNNs \cite{moons_precision} to implement EEPS \cite{eeps}, a precision-scalable processor to exploit sparsity. It was then extended into Envision \cite{envision}, with dynamic scaling of voltage, accuracy and frequency. DNPU \cite{dnpu} also supports precision scaling, but it has a much larger chip area of 16 mm$^2$. There is also some research in designing analog circuits to accelerate smaller CNNs \cite{analog_cnn}.

The architectures presented in literature have traditionally relied on scratchpads (SRAMs inside each of the hundreds of PEs), in addition to large global SRAMs, for data reuse. This typically results in over half of the overall chip area being utilized, and a significant amount of energy being consumed for scratchpads \cite{eyeriss, dsip}. Furthermore, most architectures focus on improving either performance efficiency or arithmetic intensity. Ones that report high PE utilization factors demonstrate lower overall performance efficiencies when considering their wasted clock cycles \cite{multi_mode, zasca}. Some architectures employ multiple dataflows requiring large multiplexers and idle datapaths \cite{carla,dsip}. 

This paper presents the first-generation design and implementation of the Kraken architecture and its corresponding uniform dataflow. Kraken is able to accelerate the convolutional and fully-connected layers of CNNs \cite{cnn_survey}, along with matrix products required for other DNN types such as the attention-based transformers \cite{attention} that are rising in popularity. The key contributions of this paper are as follows:

\begin{enumerate}
  \item A spatial architecture where PEs are arranged into $R{=}\vR{}$ rows and $C{=}\vC{}$ cores. Each PE consists of just a multiplier, accumulator, and a 2-way multiplexer, allowing \vNumPEs{} PEs to be packed in an area of \vArea{} \si{mm^2}.
  \item A uniform dataflow that treats fully-connected layers and matrix products as special cases of convolutional layers, outperforming the state-of-the-art in overall performance efficiency, arithmetic intensity, and memory accesses.
  \item Multiple levels of data reuse without requiring scratchpads: Output partial sums are reused within accumulators, weights are rotated in a global buffer, input activations are reused in horizontal and vertical convolutions.
  \item Elastic grouping, where $C{=}\vC{}$ cores dynamically reconfigure within one clock, using a header of just 64 bits, to process convolutional layers of different filter sizes while maintaining a high PE utilization.
  \item Decentralized control: The configuration propagates with data, reconfiguring each part on the fly, without stalling the engine, making Kraken the \emph{first accelerator} to the best of our knowledge to achieve such decentralized, dynamic reconfiguration.
  \item Detailed performance analysis: The performance efficiency, number of memory accesses, and arithmetic intensity of Kraken are derived as accurate functions of $R$ and $C$, and optimized to find the best static configuration $R{\times}C = \vR{}{\times}\vC{}$ for common DNNs.
  \item Thorough comparison of performance with prior works, benchmarked on AlexNet, VGG-16, and ResNet-50.
  \item Implementation in TSMC \vTechNm{}-nm, which outperforms the state-of-the-art \cite{carla} with \vGopsPerAreaTimesCarla{}${\times}$ more Gops/mm$^2$ and \vGopsPerWTimesCarla{}${\times}$ more Gops/W in just \vAreaTimesCarla{}${\times}$ the area.
\end{enumerate}

\section{Preliminaries}

\subsection{Convolutional Layers}

\begin{figure} 
  \centering
  \def\svgwidth{0.95\columnwidth}
  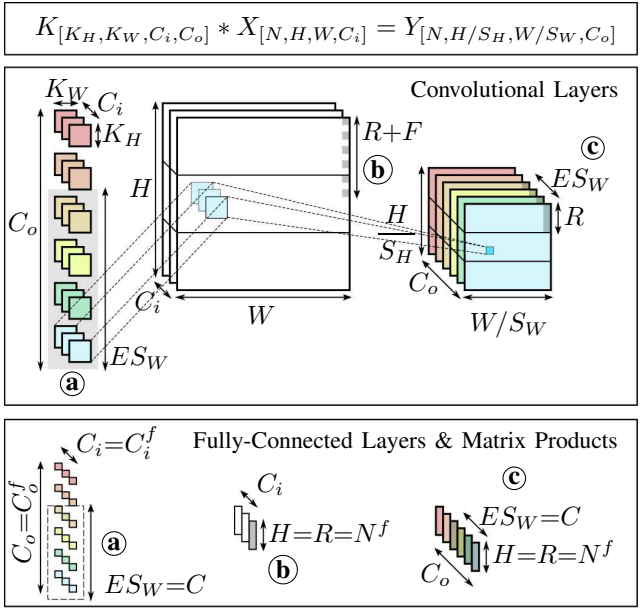
  \caption{Operations and shape parameters of convolutional layers ($N{=}1$ shown), fully-connected layers, and matrix products. Shaded pixels are stored and reused in (a) R-SRAM of weights rotator, (b) pixel shifter, and (c) PE Array, as further described in Sec. \ref{sec:df}}
  \label{fig:conv}
  \vspace{-2ex}
\end{figure}

Convolutional layers of a CNN are primarily composed of high-dimensional convolutions used for feature extraction. In such a layer, a spatial convolution followed by a depthwise dot-product between four-dimensional (4-D) arrays of input ($X_{[N,H,W,C_i]}$) and kernel weights ($K_{[K_H,K_W,C_i,C_o]}$) yield a 4-D output array ($Y_{[N,H/S_H,W/S_W,C_o]}$). The shape parameters of a convolutional layer are $N$ (batch size), $H,W$ (height and width of the input), $K_H,K_W$ (corresponding spatial dimensions of the kernel) and $C_i,C_o$ (number of input and output channels/filters). Fig. \ref{fig:conv} demonstrates the convolution operation and its shape parameters.

For the first layer, the input $X$ is composed of the batch of images, and for the subsequent layers it is made of activations of the previous layer. For each output channel $c_o$ ($=0,1,\ldots,C_o$) and input channel $c_i$, a 2-D filter of size $(K_H,K_W)$ is strided by $(S_H,S_W)$ along the $(H,W)$ dimensions of input to perform a 2-D convolution. Typically the input pixels are zero-padded to ensure the output has the same spatial dimensions but downsampled by the stride: $(H/S_H,W/S_W)$. The resulting $C_i$ number of such 2-D arrays for each output channel $c_o$ are summed together to produce a feature map. $N$ input images of a batch are processed this way, generating the output array $Y$. This operation is described as

\begin{gather}
  \label{eq:conv}
  Y_{[n,h/S_H,w/S_W,c_o]} {=} \sum^{C_i{-}1}_{c_i=0} \sum^{K_H{-}1}_{k_h=0} \sum^{K_W{-}1}_{k_w=0}  X_{[n,h^\prime,w^\prime,c_i]} K_{[k_h, k_w, c_i, c_o]}
\end{gather}
where $h^\prime = h + k_h$ and $w^\prime = w + k_w$. Lowercase terms denote the index variables that range from 0 to their uppercase counterparts, e.g., $n \in [0,N)$ and $c_o \in [0,C_o)$.

\subsection{Matrix Products and Fully-Connected Layers}

A fully-connected layer performs a matrix product between a batched 2-D input array $X_{[N^f,C_i^f]}$ and a 2-D weights array $K_{[C_i^f,C_o^f]}$ to produce the batched 2-D output array $Y_{[N^f,C_o^f]}$:
\begin{align}
  Y_{[N^f,C_o^f]} &= X_{[N^f,C_i^f]}K_{[C_i^f,C_o^f]}. \label{eq:fc}
\end{align}
Fully connected layers are used as the last few layers of a typical CNN for feature detection, while matrix products are required in the attention function of transformers, and for training of any kind of neural network. 

\subsection{Neural Network Architectures}
Neural networks are built as directed graphs of layers. While Kraken can accelerate other types of DNNs as well, in this work, we focus on CNNs. CNNs are built such that, as an image flows through a CNN, it is downsampled by integer factors in the spatial dimensions ($H,W$) due to striding and pooling. This ensures the extraction of more global features from local ones as we go deeper into the network. Meanwhile, channels typically increase from 3 at the first layer to 1024 or 2048 near the last layers to extract more complex features.

\begin{table}[t!]
  \centering
  \def\arraystretch{1.2}
\begin{threeparttable}
  \caption{CNNs considered for benchmarking.}
  \label{table:cnns}
  \begin{tabular}{c                                                                 r                                    c                                                                     c                           c                                                                              } \toprule
                                                                                  &                                    & AlexNet \cite{alexnet}                                              & VGG-16 \cite{vgg16}       & ResNet-50 \cite{resnet50}                                                     \\ \midrule
  \multicolumn{2}{r}{\STAB{Top-5 accuracy \\ on ImageNet}}                                                             & 79.06\%                                                             & 90.37\%                   & 92.93\%                                                                       \\ \midrule
  \parbox[t]{0mm}{\multirow{6}{*}{\STAB{\rotatebox[origin=c]{90}{Conv Layers}}}}  & \STAB{$(K,S){\times}$\\\# Layers}  & \STAB{ $(11,4) \times 1$\\ $(5,1)  \times 1$ \\ $(3,1)  \times 3$}  & \STAB{ $(3,1) \times 13$} & \STAB{ $(7,2) \times 1 $ \\ $(3,1) \times 16$ \\ $(1,1) \times 36$ \tnote{*} } \\ 
                                                                                  & \# $\text{MAC}_\text{w/zpad}$      & \vAlexMacsZeros{}                                                   & \vVggMacsZeros{}          & \vResMacsZeros{}                                                              \\ 
                                                                                  & \# $\text{MAC}_\text{valid}$       & \vAlexMacsValid{}                                                   & \vVggMacsValid{}          & \vResMacsValid{}                                                              \\ 
                                                                                  & $M_K$                              & 2.4 M                                                               & 14.7 M                    & 23.5 M                                                                        \\ 
                                                                                  & $M_X$                              & 299.0 K                                                             & 9.1 M                     & 8.0 M                                                                         \\ 
                                                                                  & $M_Y$                              & 650.0 K                                                             & 13.5 M                    & 10.6 M                                                                        \\ \midrule 
  \parbox[t]{0mm}{\multirow{5}{*}{\STAB{\rotatebox[origin=c]{90}{FC Layers}}}}    & \# Layers                          & 3                                                                   & 3                         & 1                                                                             \\ 
                                                                                  & \# MAC                             & \vAlexMacsFc{}                                                      & \vVggMacsFc   {}          & \vResMacsFc   {}                                                              \\ 
                                                                                  & $M_K$                              & 55.5 M                                                              & 123.6 M                   & 2.0 M                                                                         \\ 
                                                                                  & $M_X$                              & 14.3 K                                                              & 33.3 K                    & 2.0 K                                                                         \\ 
                                                                                  & $M_Y$                              & 9.2 K                                                               & 9.2 K                     & 1.0 K                                                                         \\ \bottomrule 
  \end{tabular}

  \begin{tablenotes}
    \item[*] $\dagger (K,S)=(1,2)$ layers can be processed as $(1,1)$
  \end{tablenotes}
\end{threeparttable}
\end{table}

The number of MAC operations in the $j^\text{th}$ layer can be calculated as
\begin{align}
  \text{\# MAC}_{\text{w/zpad}} &= N (H/S_H) (W/S_W) K_H K_W C_o C_i \label{eq:macs_with_zeros} \\
  \text{\# MAC}_{\text{valid}}  &= N ((H/S_H) (W/S_W) K_H K_W - Z) C_o C_i, \label{eq:macs}
\end{align}
where $H,W,C_o,K_H,K_W = 1$ for fully-connected layers.

We note that, similar to the analysis by Ahmadi \etal{} \cite{carla}, but unlike that of Chen \etal{} \cite{eyeriss} \cite{eyeriss_v2} and Ardakani \etal{} \cite{zasca}, we ignore the MAC operations corresponding to the zero paddings $(Z_j)$ when calculating the valid number of MAC operations. While this results in a lower estimate for actual performance, it better reflects the engine's capability. Furthermore, the exact number of off-chip memory accesses needed to fetch the input ($M_{X,j}$) and kernel ($M_{K,j}$), and store the outputs ($M_{Y,j}$) of the $j^\text{th}$ layer can be calculated as 
\begin{align}
  M_{X,j}   &= N H W C_i \nonumber \\
  M_{K,j}   &= K_H K_W C_i C_o \nonumber \\
  M_{Y,j}   &= N (H/S_H) (W/S_W) C_o. \nonumber
\end{align}
The number of memory accesses per layer and per inference can be calculated as their sums.

With AlexNet \cite{alexnet}, acing the ImageNet Large Scale Visual Recognition Challenge (ILSVRC) in 2012, the race to build deeper CNNs with higher accuracy began. Today, every few months, in every subfield of machine vision, a novel state-of-the-art CNN architecture is introduced to outperform the last. While any convolutional or fully-connected layer in such models is characterized by the shape parameters outlined above and can hence be accelerated by Kraken, for the purposes of benchmarking and comparison, AlexNet \cite{alexnet}, VGG-16 \cite{vgg16} and ResNet50 \cite{resnet50} are chosen in accordance with the prior works \cite{eyeriss, zasca, carla}.

\subsection{Quantization}

Integer quantization with 8-bits \cite{quant_paper} has become the industry standard for inference of DNNs. A trained network can be easily quantized (\textit{post-training quantization}) with a slight reduction of accuracy. Modern machine learning frameworks such as Tensorflow \cite{tflite} and PyTorch \cite{pytorch_quant} also widely support \textit{quantization-aware training}, where a trained network is further trained, taking quantization effects into account. This yields 8-bit inference without any noticeable degradation in accuracy for most DNNs \cite{quant_paper_2}. Bias terms ignored in equations \eqref{eq:conv} and \eqref{eq:fc} can be folded into the requantization parameters.
\section{Architecture Design}

This section describes the Kraken architecture, co-designed around its dataflow. Therefore, relevant subsections of Sec. \ref{sec:df} are referred appropriately. Kraken engine is built as a 2-D array of PEs, statically configured into $R$ \textit{rows} and $C$ \textit{cores}. Cores are elastically grouped into $E$ groups, as demonstrated in Fig. \ref{fig:system}. Each elastic group computes $S_W$ output channels, such that $R$ among $H/S_W$ rows and $ES_W$ among $C_o$ channels of the output array are calculated in parallel. The $G$ cores in each elastic group compute partial sums (further described in Sec.~\ref{subsec:df:strided}), which are shifted and accumulated $K_W$ times to the right inside the same accumulators to perform the horizontal convolution. Therefore, $E S_W R$ full output pixels are computed in parallel and released together every $1+ C_i K_H$ clocks, which are then transferred to the off-chip memory without stalling the engine. 

\subsection{Processing Element (PE)}

The simplicity of the processing element is a unique and distinguishing feature of Kraken. Traditionally accelerators are built with a large SRAM or a register file inside each of their hundreds of PEs to store partial sums, weights or inputs to make data reuse possible, as further detailed in Sec. \ref{subsec:results:compare}. Due to this complexity multiplied by the sheer number of PEs, these designs fail to pack more PEs in their chip, resulting in a fewer operations per area. The routing complexity inside such complex PEs would also reduce fmax.

\begin{figure} 
  \centering
  \def\svgwidth{0.95\columnwidth}
  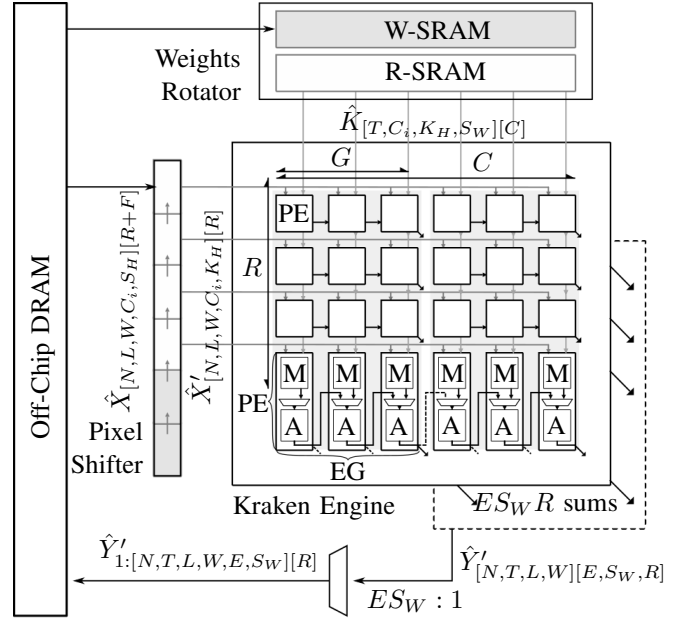
  \caption{Kraken $R{\times}C = 4{\times}6$, elastically grouped into $E{=}2$ EGs with $G{=}3$ for $(K_W,S_W){=}(3,1)$ convolution.}
  \label{fig:system}
  \vspace{-2ex}
\end{figure}

Kraken's uniform dataflow eliminates the need for SRAMs, register files, and large muxes, greatly simplifying the PE. In contrast to previously proposed designs, Kraken's PE consists of just the bare-bones: a multiplier, an accumulator with bypass, and a 2-way multiplexer (fig. \ref{fig:system}) which allows both shift-accumulation of partial sums and elastic grouping.

\subsection{Elastic Group (EG)}
For any convolutional layer, $C$ cores of the engine get \textit{elastically grouped} into $E$ EGs with $G$ cores per group, where 
\begin{align}
  G = K_W+S_W-1 \label{eq:g} \\
  E = \left\lfloor \frac{C}{G} \right\rfloor. \label{eq:e}
\end{align}
As Kraken's PE array is stateless, the multiplexers at the edges of an EG simply respond to the configuration bits tied to the wide data packets from the weights rotator to group on the fly elastically. As $W$ columns of the input array are loaded sequentially, weights are interleaved to produce partial sums corresponding to $S_W$ number of output channels in each elastic group while maintaining high utilization.

Within each EG, $G$ cores compute the partial sums of horizontal convolution. At the end of every $C_i K_H$ clocks, the multiplexers are set and partial sums are shifted into the accumulator of the core on right. The output filter extracts the results of the appropriate cores of each EG, each of which are the accumulation of $K_W C_i K_H$ products, i.e., the full output convolution sum. The interleaving of channels and strided horizontal convolution are further described in Sec. \ref{subsec:df:strided}.

During the operation, $C \% G$ cores (if any) remain idle, where \% denotes the modulo division. This number is low, as elastic groups can stretch to fill the entire span of all $C$ cores. The lack of rigid boundaries (hence elastic) between groups of cores (unlike CARLA \cite{carla} and ZASCA \cite{zasca}) while maintaining very low routing complexity enables Kraken to achieve high performance efficiency and utilization. 

\subsection{Pixel Shifter}

A small shift register bank of depth $R + \max\{F\}$ and a bank of AXI-Stream adapters (datawidth converters) make the pixel shifter. The shift factor $F$ for a given layer is defined as
\begin{equation}
  F= \lceil K_H / S_H \rceil -1. \label{eq:f}
\end{equation} 
The first $R$ registers directly supply data to the engine without any multiplexers, which helps to meet timing at high fmax. The registers are shifted $K_H$ times to enable the engine to perform strided vertical convolution optimally, as further described in Sec. \ref{subsec:v_conv}. Only the adapters needed for a given set of $(K_H,S_H)$ combinations can be instantiated during synthesis. For example, to process AlexNet, VGG-16, and ResNet-50 (Table \ref{table:cnns}), $8 \to R, R+2, R+3, R+4$ adapters are synthesized and multiplexed into the shift register bank. 

\subsection{Weights Rotator}

Two SRAMs, each $C$ words wide and $\max\{S_W C_i K_W\}$ rows deep and a 2-stage AXI Stream register pipeline to mask their latency, and an AXI Stream adapter make a weights rotator. It is worth noting that these two global buffers are the only on-chip memories in the system. Memory compilers are able to optimize large, global SRAMs and save on-chip area, in contrast to the prior works that employ hundreds of smaller per-PE SRAMs, in addition to their large global SRAM buffers.

During each iteration $t$, $C_i K_H S_W C$ kernel words required for the next iteration $t{+}1$ are slowly pre-fetched from the off-chip memory through a low-bandwidth, low-priority AXI-4 bus and filled into W-SRAM. At the end of an iteration, the two SRAMs switch their roles. The newly filled SRAM becomes R-SRAM and delivers the weights through a $C$ words wide AXI4-Stream bus to the $C$ cores of the engine. This datapath is registered at the pipeline registers, which helps to meet timing at high fmax. Inside each core, the same weight value is broadcasted to $R$ number of PEs. The weights are rotated $NLW$ times throughout the iteration, maximizing the reuse of weights to minimize DRAM accesses compared to prior works. 

\subsection{Output Pipe}

Without stalling the engine, a shift register bank of $RC$ words receives a copy of the data from the accumulators of the PE array, and shifts them along its $C$ dimension. A bank of multiplexers filter the full output sums from these $C$ groups into the $\lfloor C/3 \rfloor$ groups of a second shift register bank of depth $R \lfloor C/3 \rfloor$. The second bank shifts its $[R,ES_W]$ valid outputs into an $R$ words wide AXI4-Stream which is then sent out to DRAM.

\subsection{Static Configurability}

Kraken's implementation is highly parametrized. In addition to $R$ rows and $C$ cores of PEs, it can be synthesized for any set of word widths. The multiplier latency can be adjusted to improve timing and is set to zero in our implementation. SRAM width of weights rotator can be chosen as $\text{max}\{S_W C_i K_W\}$ for the set of CNNs that needs to be processed. The number of shift registers in the pixel shifter and output filter, AXI-Stream adapters, and the multiplexers can be synthesized either for a given set of $(K,S)$ values or for all possible combinations. 

\subsection{On-the-fly Dynamic Reconfiguration}

After being implemented in a static configuration, accelerators need to be dynamically reconfigured before processing each layer to assign work to the PEs optimally during runtime. For Kraken, headers of 64 configuration bits are pre-pended to the $\hat{X}$ (input) and $\hat{K}$ (kernel) AXI Stream packets and are streamed into the system through the datapath. In a single clock cycle, the pixel shifter and the weights rotator load the configuration bits that specify $K_H, K_W, S_H, S_W, C_i, F$ for the upcoming layer into their registers. This data, about two bytes wide, is appended to the data stream that is $R{+}C{=}103$ bytes wide. Each part of the system: multiplexers in AXI Stream adapter banks, pixel shifter, PE array, and the output shifter, each react to the configuration bits available at that point in the system, in a decentralized fashion. In the clock cycle following the completion of feeding the $\hat{X}$ and $\hat{K}$ of a layer, the configuration registers are updated with new values without stalling the engine. The modules downstream continue processing the old data and only update their behavior when the new data reaches them. This on-the-fly dynamic reconfiguration allows back-to-back processing of layers without hardware overhead and helps to achieve high fmax.
\section{Uniform Dataflow} \label{sec:df}

Dataflow is the pattern in which the 4-D arrays of input $X$, kernel $K$, and output $Y$ (each with varying shape parameters) are restructured into $\hat{X}$,$\hat{K}$, and $\hat{Y}$, respectively, and orchestrated through the fixed architecture of the PE array of $R$ rows and $C$ cores. Kraken's dataflow outperforms prior works in maximizing the reuse of data and utilizing its PEs, using a bare-bones PE architecture. This section describes Kraken's dataflow in detail, while algorithm \ref{alg:dataflow} presents a summary. 

In a nutshell, height ($H/S_w$) of the output $Y$ is split into $L$ blocks, each $R$ pixels high. The $C_o$ output channels are split into $T$ iterations, each with $ES_W$ channels. $R$ rows and $ES_W$ channels of the output are computed in parallel (fig. \ref{fig:conv}). Vertical convolution ($\Sigma^{K_H}$), depthwise dot product ($\Sigma^{C_i}$), and horizontal convolution ($\Sigma^{K_W}$) are performed in the said order to produce $RES_W$ full output pixels every $q_{kc}$ clock cycles. This is repeated over $W$ input columns, $L$ blocks, $N$ batches and $T$ iterations to produce the full output array $\hat{Y}^\prime$, where

\begin{align}
  L      &= \left\lceil \frac{H}{RS_H} \right\rceil \label{eq:l} \\
  T      &= \left\lceil \frac{C_o}{ES_W} \right\rceil \label{eq:t} \\
  q_{kc} &= 1 + K_H C_i. \label{q_kc:l}
\end{align}

Kraken's data tiling (restructuring) is expressed via a custom notation. Its order is based on the C-style array indices, also known as the row-major order, that specifies how multidimensional arrays are typically stored in a linear computer memory. In addition, two pairs of brackets show the data movement, separating the number of data beats (serial) and number of words in a data beat (parallel). For example, $\alpha{:} [1,2][3]$ denotes a 3-D C-style array of 6 elements (C: \textit{$ \langle \textit{type} \rangle $ alpha [1][2][3];}) stored in the flattened order: $\{ \alpha_{[0,0,0]}, \alpha_{[0,0,1]},\alpha_{[0,0,2]},\alpha_{[0,1,0]},\alpha_{[0,1,1]},\alpha_{[0,1,2]}\}$ and streamed through a 3-words wide port, in $2{\times}1{=}2$ clock cycles (data beats). The data orchestration is described in its loop nest representation of parallel and serial nested loops operating on these multidimensional arrays. Therefore, the number of clock cycles required to move data, the shape of data in a parallel bus, and the order in which data is moved are all expressed through the introduced notation. Kraken's dataflow is first introduced using the shape parameters of convolutional layers. Fully-connected layers and matrix products are then optimally expressed as special cases of the described dataflow. 

It should be noted that $K_j {\to} \hat{K}_j$ for all layers is performed offline and all $\hat{K}_j$ are stored in the DRAM in advance. Whereas $X_0 {\to} \hat{X}_0$ is performed once per inference for the first layer, $\hat{Y}^\prime_j {\to} \hat{Y}_j {=} \hat{X}_{j+1}$ is performed per pixel as data is streamed out of the engine, such that no clocks are wasted between layers. Therefore, the restructurings are all $\mathcal{O}(n)$ in space and time, and have \emph{no performance overhead}. 

\begin{table}[!t]
  \centering
  \def\arraystretch{1.2}

\begin{threeparttable}

  \caption{Pixel shifting for strided vertical convolution with $R,K_H,S_H = 4,7,2$}
  \label{table:clocks_sv}

  \begin{tabular}{|r|c c c c|c c c|} 
  \hline
            & \multicolumn{7}{c|}{ \# clock cycle} \\  \hline
   reg      & \cellcolor{lg} 1            & 2            & 3            & 4            & \cellcolor{lg} 5            & 6            & 7            \\ \hline
   $R_0$    & \cellcolor{lg} $x_{h_0}$    & $x_{h_2}$    & $x_{h_4}$    & $x_{h_6}$    & \cellcolor{lg} $x_{h_1}$    & $x_{h_3}$    & $x_{h_5}$    \\
   $R_1$    & \cellcolor{lg} $x_{h_2}$    & $x_{h_4}$    & $x_{h_6}$    & $x_{h_8}$    & \cellcolor{lg} $x_{h_3}$    & $x_{h_5}$    & $x_{h_7}$    \\
   $R_2$    & \cellcolor{lg} $x_{h_4}$    & $x_{h_6}$    & $x_{h_8}$    & $x_{h_{10}}$ & \cellcolor{lg} $x_{h_5}$    & $x_{h_7}$    & $x_{h_9}$    \\
   $R_3$    & \cellcolor{lg} $x_{h_6}$    & $x_{h_8}$    & $x_{h_{10}}$ & $x_{h_{12}}$ & \cellcolor{lg} $x_{h_7}$    & $x_{h_9}$    & $x_{h_{11}}$ \\ \hline
   $R_4$    & \cellcolor{lg} $x_{h_8}$    & $x_{h_{10}}$ & $x_{h_{12}}$ &              & \cellcolor{lg} $x_{h_9}$    & $x_{h_{11}}$ &              \\
   $R_5$    & \cellcolor{lg} $x_{h_{10}}$ & $x_{h_{12}}$ &              &              & \cellcolor{lg} $x_{h_{11}}$ &              &              \\
   $R_6$    & \cellcolor{lg} $x_{h_{12}}$ &              &              &              & \cellcolor{lg}              &              &              \\ \hline
  \end{tabular}

\end{threeparttable}
\end{table}

\begin{algorithm}
  \caption{Kraken's Uniform Dataflow}
  \label{alg:dataflow}
  \nonl
  \vspace{0.1cm}
  \text{Pixels in DRAM:} \nonl
  \begin{flalign*}
    X       &: [N,H,W,C_i] &\\
    X_1     &: [N,L,RS_H,W,C_i]      & \text{split} &\\
    X_2     &: [N,L,RS_H+FS_H,W,C_i] & \text{padding}&\\
    X_3     &: [N,L,R+F,S_H,W,C_i]   & \text{split}&\\
    \hat{X} &: \underbrace{[N,L,W,C_i,S_H]}_\text{data beats} \underbrace{[R+F]}_\text{parallel words} & \text{transpose} &
  \end{flalign*}
  \nonl
  \hrulefill \\
  \text{Pixels via Shifter:} \nonl
  \begin{flalign*}
    \hat{X_1}        &: [N,L,W,C_i,S_H,F^\prime][R] & \hat{X} \text{ shifted } F^\prime \text{ times} &\\
    \hat{X}^\prime &: [N,L,W,C_i,K_H][R] & \text{s.t shifted } K_H \text{ times} &
  \end{flalign*}
  \nonl
  \hrulefill \\ 
  \text{Kernel in DRAM \& via Weights Rotator:} \nonl
  \begin{flalign*}
    K       &: [K_H,K_W,C_i,C_o] &\\
    K_1     &: [K_H,K_W,C_i,T,E,S_W] & \text{split} &\\
    K_2     &: [T,C_i,K_H,E,K_W,S_W]  & \text{transpose} & \\
    K_2     &: [T,C_i,K_H,S_W][E,G]  & \text{interleave} & \\
    \hat{K} &: [T,C_i,K_H,S_W][C] &
  \end{flalign*}
  \hrulefill
  
  \SetKwBlock{DoParallel}{do in parallel}{end}
  \SetInd{0.3em}{0.3em}
  
  \BlankLine
  \text{Loop Nest Representation of Dataflow:}
  \BlankLine

  \For{$t \in [0,T)$}{ \label{alg:df:t}
    \For{$n \in [0,N)$}{   \label{alg:df:n}
      \For{$l \in [0,L)$}{      \label{alg:df:l}
        \For{$w \in [0,W)$}{    \label{alg:df:w}
        \nonl $A[R,E,G] \gets 0$ \tcp*[h]{clear all accumulators} \\
          \For{$c_i \in [0,C_i)$}{ \label{alg:df:ci}
            \For{$k_{h} \in [0,K_H)$}{ \label{alg:df:kh}
              \BlankLine
              \nonl \DoParallel{
              \BlankLine
              \For{$r \in [0,R)$}{ \label{alg:df:r}
                \For{$e \in [0,E)$}{ \label{alg:df:e}
                  \For{$g \in [0,G)$}{ \label{alg:df:g}
                  \nonl let $\hat{x}^\prime {=} \hat{X}_{[n,l,w+(g+w\%S_W)/s_w,c_i][r+k_h]}$ \\
                    \For{$s_w \in [0,S_W)$}{ \label{alg:df:sw}
                      \If{$(g+w\%S_W)\%S_W {=} s_w$}{
                        \nonl let $\hat{k} = \hat{K}_{[t,c_i,k_h,s_w][e,g]}$ \\
                        \nonl $A_{[r,e,g]} \pluseq \hat{x} \cdot \hat{k}$ \\
                        \BlankLine
                      }                        
                      \If{$g{\neq}0$ \& last $K_H, C_i$}{
                              $A_{[r,e,g-1]} \pluseq A_{[r,e,g]}$ \label{alg:df:shift} \\
                        \nonl $\hat{Y}^\prime_{[n,t,l,w][e,s_w,r]} \gets A_{[r,e,G{-}S_W{+}s_w]}$

}}}}}}}}}}}}
\nonl
\hrulefill \\ \nonl
\text{Output pipe \& DRAM storage:} \nonl
\begin{flalign*}
  \hat{Y}^\prime       &: [T,N,L,W][E,S_W,R] & \text{conv out} &\\
  \hat{Y}^\prime_1       &: [T,N,L,W,E,S_W][R] & \text{system out} &\\
  \hat{Y}^\prime_2       &: [N,L,W,T,E,S_W,R]  & \text{transpose}  &\\
  \hat{Y}^\prime_3       &: [N,L,W,C_o,R]  &\\
  \hat{Y}         &: [N,L,W,C_o,S_H][R+F] & \text{pad \& store as next } \hat{X} &
\end{flalign*}
\end{algorithm}

\subsection{Strided Vertical Convolution and Depthwise Dot Product} \label{subsec:v_conv}

$R$ rows of the PE array are tasked with computing the $R$ consecutive rows of the output $Y$. Therefore, for vertical convolution, each PE row needs to be fed with $K_H$ consecutive rows of the input $X$, while they calculate the pixels that are $S_H$ apart. As a result, while many of the same input pixels are reused between $R$ rows, they cannot be linearly shifted due to striding. Kraken's novel dataflow interleaves the pixels in memory to perform any strided vertical convolution as described below, to avoid additional registers and multiplexers required for nonlinear shifting patterns.

\newcolumntype{?}{!{\vrule width 1pt}}

\begin{table}
  \centering
  \def\arraystretch{1.2}
\begin{threeparttable}
  \caption{Dataflow of partial sums $\sigma_{w,k_w}$ inside an Elastic Group of $G{=}5$ cores when $W, K_W, S_W = 8,5,1$}
  \label{table:clocks_s1}

  \begin{tabular}{|c|c?l|p{0.08\columnwidth}|p{0.08\columnwidth}|p{0.08\columnwidth}|p{0.08\columnwidth}|p{0.08\columnwidth}|} 
  \hline
  clk \#      & $x_w$      & $g_0$           & $g_1$                          &  $g_2$                                                                           &  $g_3$                                                                                          &  $g_4$                                                                                                      \\ \hline
  $1 q_{kc}$  & $x_{w_0}$  & $\sigma_{0,0}$  & $\sigma_{0,1}$                 &  $\sigma_{0,2}$                                                                  &                                                                                                 &                                                                                                             \\ \hline
  $2 q_{kc}$  & $x_{w_1}$  & $\sigma_{1,0}$  & $\sigma_{1,1}  + \sigma_{0,0}$ &  $\sigma_{1,2}  +  \sigma_{0,1}$                                                 &  $\sigma_{1,3} + \sigma_{0,2}$                                                                  &                                                                                                             \\ \hline
  $3 q_{kc}$  & $x_{w_2}$  & $\sigma_{2,0}$  & $\sigma_{2,1}  + \sigma_{1,0}$ &  $\sigma_{2,2}  +  \sigma_{1,1}  + \sigma_{0,0}$                                 &  $\sigma_{2,3} + \sigma_{1,2}  +  \sigma_{0,1}$                                                 &  $\sigma_{2,4} + \sigma_{1,3} + \sigma_{0,2}                               = \circled{$\boldsymbol{y_0}$} $ \\ \hline
  $4 q_{kc}$  & $x_{w_3}$  & $\sigma_{3,0}$  & $\sigma_{3,1}  + \sigma_{2,0}$ &  $\sigma_{3,2}  +  \sigma_{2,1}  + \sigma_{1,0}$                                 &  $\sigma_{3,3} + \sigma_{2,2}  +  \sigma_{1,1}  + \sigma_{0,0}$                                 &  $\sigma_{3,4} + \sigma_{2,3} + \sigma_{1,2} + \sigma_{0,1}                = \circled{$\boldsymbol{y_1}$} $ \\ \hline
  $5 q_{kc}$  & $x_{w_4}$  & $\sigma_{4,0}$  & $\sigma_{4,1}  + \sigma_{3,0}$ &  $\sigma_{4,2}  +  \sigma_{3,1}  + \sigma_{2,0}$                                 &  $\sigma_{4,3} + \sigma_{3,2}  +  \sigma_{2,1}  + \sigma_{1,0}$                                 &  $\sigma_{4,4} + \sigma_{3,3} + \sigma_{2,2} + \sigma_{1,1} + \sigma_{0,0} = \circled{$\boldsymbol{y_2}$} $ \\ \hline
  $6 q_{kc}$  & $x_{w_5}$  & $\sigma_{5,0}$  & $\sigma_{5,1}  + \sigma_{4,0}$ &  $\sigma_{5,2}  +  \sigma_{4,1}  + \sigma_{3,0}$                                 &  $\sigma_{5,3} + \sigma_{4,2}  +  \sigma_{3,1}  + \sigma_{2,0}$                                 &  $\sigma_{5,4} + \sigma_{4,3} + \sigma_{3,2} + \sigma_{2,1} + \sigma_{1,0} = \circled{$\boldsymbol{y_3}$} $ \\ \hline
  $7 q_{kc}$  & $x_{w_6}$  &                 & $\sigma_{6,1}  + \sigma_{5,0}$ &  $\sigma_{6,2}  +  \sigma_{5,1}  + \sigma_{4,0}$                                 &  $\sigma_{6,3} + \sigma_{5,2}  +  \sigma_{4,1}  + \sigma_{3,0}$                                 &  $\sigma_{6,4} + \sigma_{5,3} + \sigma_{4,2} + \sigma_{3,1} + \sigma_{2,0} = \circled{$\boldsymbol{y_4}$} $ \\ \hline
  $8 q_{kc}$  & $x_{w_7}$  &                 &                                &  $\sigma_{7,2}  +  \sigma_{6,1}  + \sigma_{5,0} = \circled{$\boldsymbol{y_7}$}$  &  $\sigma_{7,3} + \sigma_{6,2}  +  \sigma_{5,1}  + \sigma_{4,0}= \circled{$\boldsymbol{y_6}$} $  &  $\sigma_{7,4} + \sigma_{6,3} + \sigma_{5,2} + \sigma_{4,1} + \sigma_{3,0} = \circled{$\boldsymbol{y_5}$} $ \\ \hline
  $9 q_{kc}$  & $x_{w_0}$  & $\sigma_{0,0}$  & $\sigma_{0,0}$                 &  $\sigma_{0,2}$                                                                  &                                                                                                 &                                                                                                             \\ \hline
\end{tabular}

\end{threeparttable}
\end{table}
\newcolumntype{?}{!{\vrule width 1pt}}

\begin{table}
  \centering
  \def\arraystretch{1.2}
\begin{threeparttable}
  \caption{Dataflow of partial sums $\sigma_{w,k_w}^{s_w}$ inside an Elastic Group of $G{=}6$ cores when $W, K_W, S_W = 8,5,2$}
  \label{table:clocks_s2}

  \begin{tabular}{|c|c?l|p{0.08\columnwidth}|p{0.08\columnwidth}|p{0.08\columnwidth}|p{0.08\columnwidth}|p{0.08\columnwidth}|} 
  \hline
  clk \#      & $x_w$      &                $g_0$             &                $g_1$                                &                $g_2$                                                &                $g_3$                                                                                                   &                $g_4$                                                                                                                  &                 $g_5$                                                                                                                  \\ \hline
  $1 q_{kc}$  & $x_{w_0}$  &                $\sigma_{0,0}^0$  & \cellcolor{lg} $\sigma_{0,0}^1$                     &                $\sigma_{0,2}^0$                                     & \cellcolor{lg} $\sigma_{0,2}^1$                                                                                        &                                                                                                                                       &  \cellcolor{lg}                                                                                                                        \\ \hline
  $2 q_{kc}$  & $x_{w_1}$  & \cellcolor{lg}                   &                $\sigma_{1,1}^0 + \sigma_{0,0}^0$    & \cellcolor{lg} $\sigma_{1,1}^1 + \sigma_{0,0}^1$                    &                $\sigma_{1,3}^0 + \sigma_{0,2}^0$                                                                       & \cellcolor{lg} $\sigma_{1,3}^1 + \sigma_{0,2}^1$                                                                                      &                                                                                                                                        \\ \hline
  $3 q_{kc}$  & $x_{w_2}$  &                $\sigma_{2,0}^0$  & \cellcolor{lg} $\sigma_{2,0}^1$                     &                $\sigma_{2,2}^0 + \sigma_{1,1}^0 + \sigma_{0,0}^0$   & \cellcolor{lg} $\sigma_{2,2}^1 + \sigma_{1,1}^1 + \sigma_{0,0}^1$                                                      &                $\sigma_{2,4}^0 + \sigma_{1,3}^0 + \sigma_{0,2}^0                                   = \circled{$\boldsymbol{y_0^0}$} $ &  \cellcolor{lg} $\sigma_{2,4}^1 + \sigma_{1,3}^1 + \sigma_{0,2}^1                                   = \circled{$\boldsymbol{y_0^1}$} $ \\ \hline
  $4 q_{kc}$  & $x_{w_3}$  & \cellcolor{lg}                   &                $\sigma_{3,1}^0 + \sigma_{2,0}^0$    & \cellcolor{lg} $\sigma_{3,1}^1 + \sigma_{2,0}^1$                    &                $\sigma_{3,3}^0 + \sigma_{2,2}^0 + \sigma_{1,1}^0 + \sigma_{0,0}^0$                                     & \cellcolor{lg} $\sigma_{3,3}^1 + \sigma_{2,2}^1 + \sigma_{1,1}^1 + \sigma_{0,0}^1$                                                    &                                                                                                                                        \\ \hline
  $5 q_{kc}$  & $x_{w_4}$  &                $\sigma_{4,0}^0$  & \cellcolor{lg} $\sigma_{3,0}^1$                     &                $\sigma_{4,2}^0 + \sigma_{3,1}^0 + \sigma_{2,0}^0$   & \cellcolor{lg} $\sigma_{4,2}^1 + \sigma_{3,1}^1 + \sigma_{2,0}^1$                                                      &                $\sigma_{4,4}^0 + \sigma_{3,3}^0 + \sigma_{2,2}^0 + \sigma_{1,1}^0 + \sigma_{0,0}^0 = \circled{$\boldsymbol{y_1^0}$} $ &  \cellcolor{lg} $\sigma_{4,4}^1 + \sigma_{3,3}^1 + \sigma_{2,2}^1 + \sigma_{1,1}^1 + \sigma_{0,0}^1 = \circled{$\boldsymbol{y_1^1}$} $ \\ \hline
  $6 q_{kc}$  & $x_{w_5}$  & \cellcolor{lg}                   &                $\sigma_{4,1}^0 + \sigma_{4,0}^0$    & \cellcolor{lg} $\sigma_{5,1}^1 + \sigma_{4,0}^1$                    &                $\sigma_{5,3}^0 + \sigma_{4,2}^0 + \sigma_{3,1}^0 + \sigma_{2,0}^0$                                     & \cellcolor{lg} $\sigma_{5,3}^1 + \sigma_{4,2}^1 + \sigma_{3,1}^1 + \sigma_{2,0}^1                                                   $ &                                                                                                                                        \\ \hline
  $7 q_{kc}$  & $x_{w_6}$  &                                  & \cellcolor{lg}                                      &                $\sigma_{6,2}^0 + \sigma_{5,1}^0 + \sigma_{4,0}^0$   & \cellcolor{lg} $\sigma_{6,2}^1 + \sigma_{5,1}^1 + \sigma_{4,0}^1$                                                      &                $\sigma_{6,4}^0 + \sigma_{5,3}^0 + \sigma_{4,2}^0 + \sigma_{3,1}^0 + \sigma_{2,0}^0 = \circled{$\boldsymbol{y_2^0}$} $ &  \cellcolor{lg} $\sigma_{6,4}^1 + \sigma_{5,3}^1 + \sigma_{4,2}^1 + \sigma_{3,1}^1 + \sigma_{2,0}^1 = \circled{$\boldsymbol{y_2^1}$} $ \\ \hline
  $8 q_{kc}$  & $x_{w_7}$  & \cellcolor{lg}                   &                                                     & \cellcolor{lg}                                                      &                $\sigma_{7,3}^0 + \sigma_{6,2}^0 + \sigma_{5,1}^0 + \sigma_{4,0}^0 = \circled{$\boldsymbol{y_3^0}$} $   & \cellcolor{lg} $\sigma_{7,3}^1 + \sigma_{6,2}^1 + \sigma_{5,1}^1 + \sigma_{4,0}^1                  = \circled{$\boldsymbol{y_3^1}$} $ &                                                                                                                                        \\ \hline
  $9 q_{kc}$  & $x_{w_0}$  &                $\sigma_{0,0}^0$  & \cellcolor{lg} $\sigma_{0,0}^1$                     &                $\sigma_{0,2}^0$                                     & \cellcolor{lg} $\sigma_{0,2}^1$                                                                                        &                                                                                                                                       &  \cellcolor{lg}                                                                                                                        \\ \hline
  \end{tabular}
\end{threeparttable}
\end{table}

The 4-D array of input pixels $X$ is first sliced along $H$ dimension into $L$ blocks, where each block has a height $RS_H$, producing $X_1$. Each block $l$ is then padded with $(K_H-1)/2$ bottom rows of the previous block $l{-}1$ and $F-(K_H-1)/2$ \eqref{eq:f} top rows of the next block $l{+}1$ to produce $X_2$. The height of each block is then reshaped into $[R+F,S_H]$ to produce $X_3$. Finally, the entire array is transposed into $\hat{X}$ and stored in the off-chip DRAM. This multidimensional transpose operation results in pixel interleaving as demonstrated in Table \ref{table:clocks_sv}.

The tiled input $\hat{X}$ is pulled from the DRAM into $R+F$ parallel words. $S_H$ such data beats are loaded sequentially into a small shift register bank of $R+F$ words, as demonstrated in Table \ref{table:clocks_sv}. After each such load (shaded clock cycles), the registers are shifted $F^\prime$ times, resulting in $\hat{X}_1$, where
\begin{equation}
  F^\prime =
  \begin{cases}
    \lfloor K_H / S_H \rfloor & \text{on } S_H^\text{th} \text{ (last) load} \\
    F & \text{other loads}.
  \end{cases}
  \label{eq:f_prime}
\end{equation} 
First interleaving the pixels and then shifting them $F^\prime$ times ($\hat{X}$) is equivalent to loading $K_H$ consecutive pixels into each of first $R$ registers, just in a different order ($\hat{X^\prime}$). The $R$ registers are directly connected to the $R$ rows of the PE array, ensuring each row gets the input pixels needed to calculate its strided output pixel, as demonstrated in Table \ref{table:clocks_sv}. This exploitation of data reuse in the $H$ dimension of input $X$ results in an $(F^\prime{+}1){\times}$ \emph{reduction of DRAM accesses} on the input side of pixel shifter and fewer engine stalls.

Synchronized with this shifting, the weights rotator supplies $K_H$ kernel words to the $C$ cores, allowing the PE array to first perform the strided vertical convolution, corresponding to loop \ref{alg:df:kh} in algorithm \ref{alg:dataflow}. This operation is repeated over $C_i$ input channels (loop \ref{alg:df:ci}) as the PEs accumulate the depthwise dot-product. Since input channels vary widely in depth across layers and are not shared with neighboring pixels, serially processing them allows 100\% utilization across this dimension. After the end of this operation, which takes $q_{kc}$ clocks, the pixel shifter repeats it over the next input column ($W$ dimension).

\subsection{Unstrided Horizontal Convolution (\texorpdfstring{$S_W{=}1$})}

Table \ref{table:clocks_s1} demonstrates the horizontal convolution for a simplified example, where $W,K_W{=}8,5$, $~C,G{=}5$ and $S_W,C_o,E{=}1$. The partial sums resulting from vertical convolution followed by depthwise dot product are denoted as 
\begin{align}
  \sigma_{w,k_w} = \sum^{C_i} \sum^{K_H} X_{[..,w,..]}K_{[..,k_w,..]}.
\end{align}
Here, $C{=}5$ cores get elastically grouped into $E{=}1$ elastic groups of $G{=}K_W{=}5$ cores each. At the end of each depthwise sum (loop \ref{alg:df:ci}), during a single clock cycle, the muxes in PEs are engaged. PEs of each core, except the first, receive the partial sums $\sigma_{w,k_w}$ from those of the core on the left and accumulate with their own sums.

As demonstrated, after shifting and accumulating $\lfloor K_W/2 \rfloor {=}2$ times, the last core of the EG contains the first valid output column with implicit zero padding: \smallcircled{$y_0$}. Then, in each consecutive such cycle of $q_{kc}$ clocks, the last core releases the subsequent output column \smallcircled{$y_w$}. At the last such cycle, the last $\lceil K_W/2 \rceil{=}3$ valid output columns are released in the same clock, with implicit zero paddings. Therefore, exactly $Wq_{kc}{=}8q_{kc}$ clock cycles are required to compute $W{=}8$ output columns. This operation allows Kraken to perform horizontal zero padding \emph{without extra circuitry or extra data fetches}. Accumulators flush their registers with new products from multipliers and start processing the next block, on the following clock itself.

\subsection{Strided Horizontal Convolution (any \texorpdfstring{$S_W$})}
\label{subsec:df:strided}

Equation \eqref{eq:conv} shows that horizontally strided convolution is equivalent to discarding all but one column in each stride of $S_W$ after a regular unstrided convolution. This implies that calculations along $S_W{-}1$ number of diagonals in Table \ref{table:clocks_s1} are unnecessary when striding. $S_W{-}1$ additional output channels are hence calculated through those diagonals, achieving maximal utilization using the same uniform dataflow. 

Table \ref{table:clocks_s2} demonstrates the strided horizontal convolution for a simplified example where $W,K_W {=} 8,5$, $~C,G{=}6$, $~S_W,C_o{=}2$ and $E{=}1$. The partial sums after vertical convolution and depthwise dot product are denoted as
\begin{align}
  \sigma_{w,k_w}^{s_w{=}c_o} = \sum^{C_i} \sum^{K_H} X_{[..,w,..]}K_{[..,k_w,..,c_o,..]}.
\end{align}
Generalized for any convolution, $C{=}6$ cores get elastically grouped into $E{=}1$ elastic groups, each with $G{=}6$ cores, computing $S_W{=}2$ output channels in parallel, such that all output channels are computed in $T$ iterations. The data movement in this horizontally strided convolution is demonstrated in Table \ref{table:clocks_s2}. The output channel corresponding to each partial sum is denoted by its superscript and their cells are differently shaded to clearly demonstrate the channel interleaving. $S_W{=}2$ adjacent cores perform identical computations on $S_W{=}2$ output channels. Their results \smallcircled{$y_w^{s_w}$} are released in parallel at the same clock cycles. Therefore, after every  $q_{kc}$ clocks, each row of the PE array releases $ES_W$ full output pixels, such that $ES_WR$ pixels are released by the engine. Since initial layers have a bigger filter size $K_W$ (hence smaller $E$) and latter layers have more input channels $C_i$, the data can be streamed out into the DRAM without stalling the engine. At the end of a layer, the $C$ cores would get dynamically regrouped into a new set of elastic groups without pausing their operation.

\subsection{Matrix Product and Fully-Connected Layers}

The multiplication between two matrices 
\begin{equation}
  M_{1:[H,C_i]} M_{2:[C_i,C_o]} = M_{3:[H,C_o]} \label{eq:df:matmul}
\end{equation}

is a special case of the described dataflow, where $N,W,K_H,K_W,S_H,S_W = 1$. Kraken's PE array of size $(R,C)$ computes the full submatrix $M_{3:[R,C]}$ in $C_i$ clocks and releases it without any shifting. In $TL$ such iterations, all submatrices of $M_3$ are computed. Consequently, the inference of a fully-connected layer described in \eqref{eq:fc} can be performed with $N,H,C_i,C_o = 1,N^f, C_i^f, C_o^f$. Inference batch size for the fully-connected layers $(H{=}N^f)$ can be hence chosen as $R$ to fully utilize the rows of the PE array and reduce the number of memory accesses by reusing the weights.

\subsection{Stationary-ness}

Dataflows are categorized by the type of data reuse they prioritize\cite{dnn_book}. Kraken's dataflow primarily prioritizes computing $ES_WR$ complete output pixels inside accumulators (reuse (c) in fig. \ref{fig:conv}) to simplify the PEs to their bare-bones by eliminating SRAMs and register files. This makes it output-stationary with respect to the engine. Besides, maximizing data reuse of the kernel array $K$ is of paramount importance as it is responsible for 73\% to 96\% of all data movement (see Table \ref{table:cnns}). Hence, Kraken is also designed to be weight-stationary with respect to the system. Primarily weight-stationary architectures hold their weights in register files inside their PEs. Avoiding that, Kraken prefetches weights into global SRAMs (reuse (a) in fig. \ref{fig:conv}) and rotates them thousands of times, maximizing their reuse throughout an iteration. Pixel shifting exploits data reuse in the $H$ dimension (reuse (b) in fig. \ref{fig:conv}), and periodic shift-accumulate within an elastic group exploits data reuse in the $W$ dimension to further lower the input bandwidth requirement. Therefore, Kraken is built to \emph{maximally exploit the reuse of all outputs, weights and inputs}.

\section{Performance Analysis}
\label{sec:analysis}

This section presents a detailed performance analysis of Kraken for any DNN, deriving the key metrics as exact functions, which are later optimized over a set of CNNs to find the best static configuration.

\subsection{Clock Cycles (Q)}

When processing convolutional layers with $K \neq 1$, after accumulating every $C_i K_H$ products, the multipliers pause for one clock to allow shifted accumulation. In such layers, the one clock needed to load the configuration bits does not stall the engine as the pixel shifter reduces the necessary bandwidth on the input side. When processing convolutional layers with $K_W = 1$, fully-connected layers, and matrix products, there is no pause for shifting; however, the dataflow is stalled for one clock for configuration, i.e.,
\begin{align}
  q_{s} &= \begin{cases}
  1 & \text{if (conv \& } K_W{\neq}1 \text{)} \\
  0 & \text{otherwise,}
  \end{cases} \label{eq:iota_s} \\
  q_{c} &= \begin{cases}
  0 & \text{if (conv \& } K_W{\neq}1 \text{)} \\
  1 & \text{otherwise.} 
  \end{cases}
\end{align}
Therefore, the number of clocks required for the $j^{\text{th}}$ layer is 
\begin{equation}
  Q_j = T(q_c + N L W (q_s + C_i K_H)). \label{eq:qj}
\end{equation}

\subsection{Performance Efficiency \texorpdfstring{$(\mathcal{E})$}{subseceff} }

In order to evaluate the ability of a dataflow to utilize the processing elements over the entire operation, \textit{Performance Efficiency} over a DNN or a set of DNNs is defined as
\begin{align}
  \text{Performance Efficiency} (\mathcal{E}) &= \dfrac{\text{Average Gops}}{\text{Peak Gops}} \nonumber\\
                                                &= \dfrac{\text{Valid Gops of layer}}{\text{Peak Gops of the PE array}} \nonumber\\
                                                &= \dfrac{\text{ \# MAC}_{\text{valid}}}{\text{\# PEs} \times Q} 
                                                = \dfrac{\Sigma\mathcal{E}_j Q_j}{\Sigma Q_j},
  \label{eq:p_eff}
\end{align}

where $\mathcal{E}$ of of the $j^\text{th}$ layer of a DNN is
\begin{gather}
  \mathcal{E_j} = \dfrac{\text{ \# MAC}_{\text{valid},j}}{\text{\# PEs} \times Q_j}. \nonumber
\end{gather}
It is worth mentioning that unlike in prior works \cite{eyeriss}, \cite{zasca}, only the operations that exclude zero padding are considered valid, while all clock cycles $Q_j$, including those required for reconfiguration, are considered for realistic analysis.

Note that, for fully-connected layers and matrix products, $H,C_i= N^f,C_i^f$ and $N,W,K_H,{=}1$. Hence, using \eqref{eq:macs}, \eqref{eq:l}, \eqref{eq:t}, and \eqref{eq:qj}, the performance efficiency $\mathcal{E}_j$ of Kraken over a layer can be derived as a function of the static configuration parameters $R,C$ as
\begin{align}
  \mathcal{E}_j (R,C) &= \dfrac{(N K_H K_W H W /(S_H S_W){-}Z) C_o C_i}{RCT(q_c + NLW(q_s + C_iK_H))}.
  \label{eq:kraken_ej}
\end{align}

In order to easily observe the key factors affecting $\mathcal{E}_j$, shifting and configuration clock cycles ($q_s, q_c$) can be neglected to yield the simplified function:
\begin{align}
  \mathcal{E}_j (R,C) &= \dfrac{\left( \dfrac{H}{R S_H} \right)}{ \left\lceil \dfrac{H}{R S_H} \right\rceil } \cdot
  \dfrac{C_o K_W}{C S_W \left\lceil \dfrac{C_o}{S_W  \left\lfloor \dfrac{C}{K_W + S_W-1 } \right\rfloor } \right\rceil}. \nonumber
\end{align}
Since $H$ of the layers decrease by integer factors as we progress through a CNN due to pooling and striding, $R$ can be chosen such that $H$ is evenly divisible by $RS_H$ for all layers. Observing that all but the first couple of layers of a CNN have $S_W{=}1$ and $K_W{=}3,1$, $C$ can be chosen as a multiple of 3 (as with the implemented Kraken $\vR{}{\times}\vC{}$ configuration) improving their efficiency into
\begin{align}
  \mathcal{E}_{(j>0,1)} &= \dfrac{\left( \dfrac{C_o K_W}{C} \right)}{\left\lceil \dfrac{C_o K_W}{C} \right\rceil}. \nonumber
\end{align}

\subsection{Memory Accesses (M)}

The number of memory accesses ($\hat{M}_j$) Kraken requires to compute the $j^\text{th}$ layer is the sum of the data moved as input pixels $M_{\hat{X},j}$, weights $M_{\hat{K},j}$ and output pixels $M_{\hat{Y},j}$, which are also functions of the static configuration parameters $R,C$. The total memory accesses $\hat{M}(R,C)$ can be computed as
\begin{align}
  \hat{M}(R,C) &= \Sigma \hat{M}_{j}(R,C), \label{eq:m}
\end{align}
where
\begin{align}
  \hat{M}_{j}   (R,C) &= M_{\hat{X},j}(R,C) + M_{\hat{K},j}(R,C) + M_{\hat{Y},j}(R,C) \label{eq:m_j} \nonumber \\
  M_{\hat{X},j} (R,C) &= T  N  L  W  C_i S_H (R + F) \nonumber \\
  M_{\hat{K},j} (R,C) &= T  C_i  K_H  S_W C \nonumber \\
  M_{\hat{Y},j} (R,C) &= T  N  L  W  E S_W  R. \nonumber
\end{align}

\subsection{Arithmetic Intensity (AI)}

To measure the degree of data reuse facilitated by the dataflow throughout a CNN, $\text{AI}$ is defined as
\begin{align}
  \text{AI} = \dfrac{\text{\# Valid Operations}}{\# \text{Memory Accesses}}.
\end{align}
For Kraken, $\text{AI} (R,C)$ can be computed using \eqref{eq:macs} and \eqref{eq:m} as
\begin{align}
  \text{AI} (R,C) = \dfrac{2 \times \text{\# MAC}_\text{valid}}{\hat{M}(R,C)}. \label{eq:ai_j}
\end{align}

\subsection{Memory Bandwidth Requirement}

The pixel shifter of Kraken requires $R+F$ words of input pixels in every $F^\prime$ clocks. Over iteration $t$, the weights rotator loads $C_i K_H S_W C$ number of words of the weights for the next iteration $t{+}1$. Furthermore, the output pipe needs to stream $E S_W R$ words of the previous output column $w{-}1$ within $S_W(C_i K_H {+} q_s)$ clocks, before the current output column $j$ is generated by the PE array. Therefore, the bandwidth (words/s) requirement of the input $\hat{X}$, kernel $\hat{K}$, and the output $\hat{Y}$ at frequency $f$ are computed as
\begin{align}
  \text{Bandwidth}_{\hat{X}} &= f (R+F)/F^\prime \label{eq:bw_x} \\
  \text{Bandwidth}_{\hat{K}} &= f_t \dfrac{[C_i K_H S_W C]_{(t+1)}}{[q_c + N L W (q_s + C_i K_H)]_t} \label{eq:bw_k}  \\
  \text{Bandwidth}_{\hat{Y}} &= f_w \dfrac{[E S_W R]_{(w-1)}}{[C_i K_H + q_s]_w}. \label{eq:bw_y}
\end{align}
For fully-connected layers and matrix products, take $C_i{=}C_i^f$, $C_o{=}C_o^f$,  $~F,F^\prime,q_s {=}0$, and $q_c, K_H, S_W, N, L, W, E {=} 1$.

\section{Results and Discussion}

\newcolumntype{P}[1]{>{\centering\arraybackslash}p{#1}}

\begin{table*}[ht]
  \centering
  \def\arraystretch{1.2}
\begin{threeparttable}
  \caption{Comparison with state-of-the-art implementations on Convolutional Layers}
  \label{table:compare_conv}

  \begin{tabular}{
    @{}l                              c                           c                             c                         c                         c                           c                        c                                    c                       c                     c                                                 @{}} \toprule[1.5pt] 
                                     & \multicolumn{2}{c}{JSSC’17\cite{eyeriss}   }            & \multicolumn{3}{c}{TCOMP'20\cite{zasca}}                                        & \multicolumn{2}{c}{TCAS'21\cite{carla}}                     & \multicolumn{3}{c}{ This work}                                                                \\                            
                                     & \multicolumn{2}{c}{Eyeriss                 }            & \multicolumn{3}{c}{ZASCAD            }                                          & \multicolumn{2}{c}{CARLA}                                   & \multicolumn{3}{c}{Kraken $\vR{}{\times}\vC{}$}                                               \\     \midrule                        
    Technology                       & \multicolumn{2}{c}{TSMC 65nm   }                        & \multicolumn{3}{c}{TSMC 65nm }                                                  & \multicolumn{2}{c}{TSMC 65nm      }                         & \multicolumn{3}{c}{ TSMC \vTechNm{}nm }                                                       \\
    Methodology                      & \multicolumn{2}{c}{Silicon}                             & \multicolumn{3}{c}{Place \& Route }                                             & \multicolumn{2}{c}{Synthesis }                              & \multicolumn{3}{c}{ Synthesis}                                                                \\
    \#PEs                            & \multicolumn{2}{c}{168    }                             & \multicolumn{3}{c}{192  }                                                       & \multicolumn{2}{c}{196       }                              & \multicolumn{3}{c}{ \vNumPEs{}}                                                               \\
    On-chip RAM (KB)                 & \multicolumn{2}{c}{181.5  }                             & \multicolumn{3}{c}{36.9 }                                                       & \multicolumn{2}{c}{85.5      }                              & \multicolumn{3}{c}{ \vRam{}  }                                                                \\
    Core Area (mm$^2$)               & \multicolumn{2}{c}{12.25  }                             & \multicolumn{3}{c}{6    }                                                       & \multicolumn{2}{c}{6.2       }                              & \multicolumn{3}{c}{ \vArea{} }                                                                \\
    Frequency (MHz)                  & \multicolumn{2}{c}{200    }                             & \multicolumn{3}{c}{200  }                                                       & \multicolumn{2}{c}{200       }                              & \multicolumn{3}{c}{ \vConvFreqMhz{}}                                                          \\
    Bit precision                    & \multicolumn{2}{c}{16     }                             & \multicolumn{3}{c}{16   }                                                       & \multicolumn{2}{c}{16        }                              & \multicolumn{3}{c}{ 8        }                                                                \\      \cmidrule(lr){2-3} \cmidrule(lr){4-6}  \cmidrule(lr){7-8} \cmidrule(l){9-11} 
                                     & AlexNet                    & VGG16                      & AlexNet                   & VGG16                    & ResNet50                 & VGG16                   & ResNet50                          & AlexNet                       & VGG16                         & ResNet50                      \\      \cmidrule(lr){2-3} \cmidrule(lr){4-6}  \cmidrule(lr){7-8} \cmidrule(l){9-11} 
    Performance Efficiency (\%)      & 63.6                       & 30.8                       & \vAlexZascadConvPEf    {} & \vVggZascadConvPEf    {} & \vResZascadConvPEf    {} & \vVggCarlaConvPEf    {} & \textbf{\vResCarlaConvPEf    {}}  & \textbf{\vAlexConvPEf    {}}  & \textbf{\vVggConvPEf    {}}   &        {\vResConvPEf    {}}   \\
    Throughput (fps)                 & 34.7                       & 0.7                        & \vAlexZascadConvFps    {} & \vVggZascadConvFps    {} & \vResZascadConvFps    {} & \vVggCarlaConvFps    {} &         \vResCarlaConvFps    {}   & \textbf{\vAlexConvFps    {}}  & \textbf{\vVggConvFps    {}}   & \textbf{\vResConvFps    {}}   \\
    Latency (ms)                     & 115.3                      & 4309.5                     & \vAlexZascadConvLatency{} & \vVggZascadConvLatency{} & \vResZascadConvLatency{} & \vVggCarlaConvLatency{} &         \vResCarlaConvLatency{}   & \textbf{\vAlexConvLatency{}}  & \textbf{\vVggConvLatency{}}   & \textbf{\vResConvLatency{}}   \\
    Power (mW)                       & 278                        & 236                        & \vAlexZascadConvPower  {} & \vVggZascadConvPower  {} & \vResZascadConvPower  {} & \vVggCarlaConvPower  {} &         \vResCarlaConvPower  {}   &        {\vAlexConvPower  {}}  &        {\vVggConvPower  {}}   &        {\vResConvPower  {}}   \\      
    Batch size                       & 4                          & 3                          & \vAlexZascadConvBatch  {} & \vVggZascadConvBatch  {} & \vResZascadConvBatch  {} & \vVggCarlaConvBatch  {} &         \vResCarlaConvBatch  {}   &        {\vAlexConvBatch  {}}  &        {\vVggConvBatch  {}}   &        {\vResConvBatch  {}}   \\     
    Performance (Gops)               & 42.8                       & 20.7                       & \vAlexZascadConvGops   {} & \vVggZascadConvGops   {} & \vResZascadConvGops   {} & \vVggCarlaConvGops   {} &         \vResCarlaConvGops   {}   & \textbf{\vAlexConvGops   {}}  & \textbf{\vVggConvGops   {}}   & \textbf{\vResConvGops   {}}   \\
    Performance/Area (Gops/\si{mm^2})& 3.5                        & 1.7                        & \vAlexZascadConvGopsAr {} & \vVggZascadConvGopsAr {} & \vResZascadConvGopsAr {} & \vVggCarlaConvGopsAr {} &         \vResCarlaConvGopsAr {}   & \textbf{\vAlexConvGopsAr {}}  & \textbf{\vVggConvGopsAr {}}   & \textbf{\vResConvGopsAr {}}   \\
    Energy Efficiency (Gops/W)       & 153.8                      & 87.6                       & \vAlexZascadConvEnEf   {} & \vVggZascadConvEnEf   {} & \vResZascadConvEnEf   {} & \vVggCarlaConvEnEf   {} &         \vResCarlaConvEnEf   {}   & \textbf{\vAlexConvEnEf   {}}  & \textbf{\vVggConvEnEf   {}}   & \textbf{\vResConvEnEf   {}}   \\
    Memory Access / frame ($10^6$)   & \textbf{2.0  }             & \textbf{56.1 }             & \vAlexZascadConvMAMn   {} & \vVggZascadConvMAMn   {} & \vResZascadConvMAMn   {} & \vVggCarlaConvMAMn   {} &         \vResCarlaConvMAMn   {}   &        {\vAlexConvMAMn   {}}  &        {\vVggConvMAMn   {}}   &        {\vResConvMAMn   {}}   \\     
    Memory Access / frame (MB)       & \textbf{3.85 }             & \textbf{107.0}             & \vAlexZascadConvMAMb   {} & \vVggZascadConvMAMb   {} & \vResZascadConvMAMb   {} & \vVggCarlaConvMAMb   {} &         \vResCarlaConvMAMb   {}   &        {\vAlexConvMAMb   {}}  &        {\vVggConvMAMb   {}}   &        {\vResConvMAMb   {}}   \\     
    Arithmetic Intensity (Op/MA)     & \textbf{610.6}             & \textbf{529.1}             & \vAlexZascadConvAI     {} & \vVggZascadConvAI     {} & \vResZascadConvAI     {} & \vVggCarlaConvAI     {} &         \vResCarlaConvAI     {}   &        {\vAlexConvAI     {}}  &        {\vVggConvAI     {}}   &        {\vResConvAI     {}}   \\      \bottomrule[1.5pt]
  \end{tabular}
\end{threeparttable}
\end{table*}

\begin{table}[ht]
  \centering
  \def\arraystretch{1.2}
\begin{threeparttable}
  \caption{\textsc{Comparison with state-of-the-art implementations on Fully-connected Layers}}
  \label{table:compare_fc}
  \begin{tabular}{
    @{}p{0.23\columnwidth}           P{0.07\columnwidth}   P{0.06\columnwidth}         P{0.1\columnwidth}        P{0.07\columnwidth}             P{0.06\columnwidth}             P{0.12\columnwidth}            @{}} \toprule[1.5pt] 
                                    & \multicolumn{3}{c}{ZASCAD\cite{zasca}}                                     & \multicolumn{3}{c}{Kraken $\vR{}{\times}\vC{}$}                                               \\   \midrule                        
    Frequency (MHz)                 & \multicolumn{3}{c}{40   }                                                  & \multicolumn{3}{c}{ \vFcFreqMhz{}  }                                                          \\
                                    & AlexNet                 & VGG16                  & ResNet50                & AlexNet                       & VGG16                         & ResNet50                      \\   \cmidrule(l){2-4} \cmidrule(l){5-7} 
    Perf. Eff. (\%)                 & \vAlexZascadFcPEf    {} & \vVggZascadFcPEf    {} & \vResZascadFcPEf    {}  & \textbf{\vAlexFcPEf    {}}    & \textbf{\vVggFcPEf    {}}     & \textbf{\vResFcPEf    {}}     \\
    Throughput (fps)                & \vAlexZascadFcFps    {} & \vVggZascadFcFps    {} & \vResZascadFcFps    {}  & \textbf{\vAlexFcFps    {}}    & \textbf{\vVggFcFps    {}}     & \textbf{\vResFcFps    {}}     \\
    Latency (ms)                    & \vAlexZascadFcLatency{} & \vVggZascadFcLatency{} & \vResZascadFcLatency{}  & \textbf{\vAlexFcLatency{}}    & \textbf{\vVggFcLatency{}}     & \textbf{\vResFcLatency{}}     \\
    Power (mW)                      & \vAlexZascadFcPower  {} & \vVggZascadFcPower  {} & \vResZascadFcPower  {}  &        {\vAlexFcPower  {}}    &        {\vVggFcPower  {}}     &        {\vResFcPower  {}}     \\      
    Batch size                      & \vAlexZascadFcBatch  {} & \vVggZascadFcBatch  {} & \vResZascadFcBatch  {}  &        {\vAlexFcBatch  {}}    &        {\vVggFcBatch  {}}     &        {\vResFcBatch  {}}     \\     
    Perf. (Gops)                    & \vAlexZascadFcGops   {} & \vVggZascadFcGops   {} & \vResZascadFcGops   {}  & \textbf{\vAlexFcGops   {}}    & \textbf{\vVggFcGops   {}}     & \textbf{\vResFcGops   {}}     \\
    Gops/\si{mm^2}                  & \vAlexZascadFcGopsAr {} & \vVggZascadFcGopsAr {} & \vResZascadFcGopsAr {}  & \textbf{\vAlexFcGopsAr {}}    & \textbf{\vVggFcGopsAr {}}     & \textbf{\vResFcGopsAr {}}     \\
    En.Eff. (Gops/W)                & \vAlexZascadFcEnEf   {} & \vVggZascadFcEnEf   {} & \vResZascadFcEnEf   {}  & \textbf{\vAlexFcEnEf   {}}    & \textbf{\vVggFcEnEf   {}}     & \textbf{\vResFcEnEf   {}}     \\
    MA/frame ($10^6$)               & \vAlexZascadFcMAMn   {} & \vVggZascadFcMAMn   {} & \vResZascadFcMAMn   {}  & \textbf{\vAlexFcMAMn   {}}    & \textbf{\vVggFcMAMn   {}}     & \textbf{\vResFcMAMn   {}}     \\     
    MA/frame (MB)                   & \vAlexZascadFcMAMb   {} & \vVggZascadFcMAMb   {} & \vResZascadFcMAMb   {}  & \textbf{\vAlexFcMAMb   {}}    & \textbf{\vVggFcMAMb   {}}     & \textbf{\vResFcMAMb   {}}     \\     
    AI (Op/MA)                      & \vAlexZascadFcAI     {} & \vVggZascadFcAI     {} & \vResZascadFcAI     {}  & \textbf{\vAlexFcAI     {}}    & \textbf{\vVggFcAI     {}}     & \textbf{\vResFcAI     {}}     \\   \bottomrule[1.5pt]
  \end{tabular} 
\end{threeparttable}
\end{table}

In this section, the implementation of Kraken is described and then compared with the prior works: Eyeriss \cite{eyeriss}, MMIE/ZASCAD \cite{zasca}, and CARLA \cite{carla}. Whereas Kraken can accelerate any DNN with convolutional, fully-connected layers and matrix products, it is benchmarked on AlexNet, VGG-16, and ResNet-50 for comparison.

\subsection{Implementation}

Based on the performance analysis presented Sec. \ref{sec:analysis}, and optimizing with respect to the performance efficiency in \eqref{eq:kraken_ej} and the memory accesses in \eqref{eq:m} over the three CNNs, the static configuration that minimizes the memory accesses with overall optimal performance efficiency is calculated as $R{\times}C{=}\vR{}{\times}\vC{}$. Although slightly higher performance efficiencies can be achieved by reducing $C$ at $R{\times}C = 7{\times}15, 7{\times}24$ \& $14{\times}24$, these improvements are found to be minimal, at the expense of a much higher number of memory accesses.

The architecture of Kraken was described and verified primarily in SystemVerilog. Interfaces were implemented to comply with the stream and memory-mapped protocols from the industry-standard system bus family of ARM Advanced eXtensible Interface (AXI). After hardware verification on the FPGA: Xilinx Z-7045 Programmable SoC, the design was appropriately modified for ASIC and was synthesized using Cadence Genus with TSMC \vTechNm{}-nm GP CMOS technology. SRAMs generated using Arm Artisan Memory Compiler were packed into banks that are $\text{max}\{S_W C_i K_W\} {=}2048$ rows deep and $C{=}96$ words wide. Open-source IPs \cite{verilog_axis} were used for AXI protocol conversion.

As per \eqref{eq:bw_x}, \eqref{eq:bw_k}, and \eqref{eq:bw_y}, the peak bandwidth required for Kraken $\vR{}{\times}\vC{}$ is 26 bytes/clock for the convolutional layers (layer 1 of VGG-16) and 104 bytes/clock for the fully-connected layers. LPDDR4 memory packages offer bandwidths up to 25.6 GB/s (3200 mega transfers per second over a 64-bit IO bus) \cite{lpddr4}. Therefore, to operate well within this bandwidth, Kraken is implemented to be run at a frequency of \vConvFreqMhz{} MHz for convolutional layers and \vFcFreqMhz{} MHz for fully-connected layers.

\subsection{Comparison with State-of-the-Art Implementations}
\label{subsec:results:compare}

Table \ref{table:compare_conv} compares the results of Kraken $\vR{}{\times}\vC{}$ with the prior works on convolutional layers of AlexNet, VGG-16, and ResNet-50. Since only ZASCAD reports performance on fully-connected layers, Table \ref{table:compare_fc} compares Kraken $\vR{}{\times}\vC{}$ and ZASCAD on them.


\begin{figure}
  \centering
  \def\svgwidth{\columnwidth}
  \begin{center}
    \includegraphics{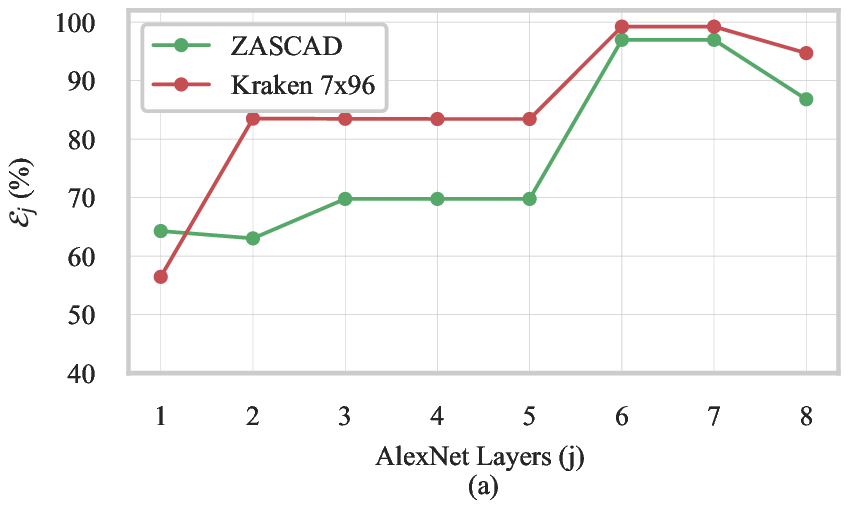}
    \includegraphics{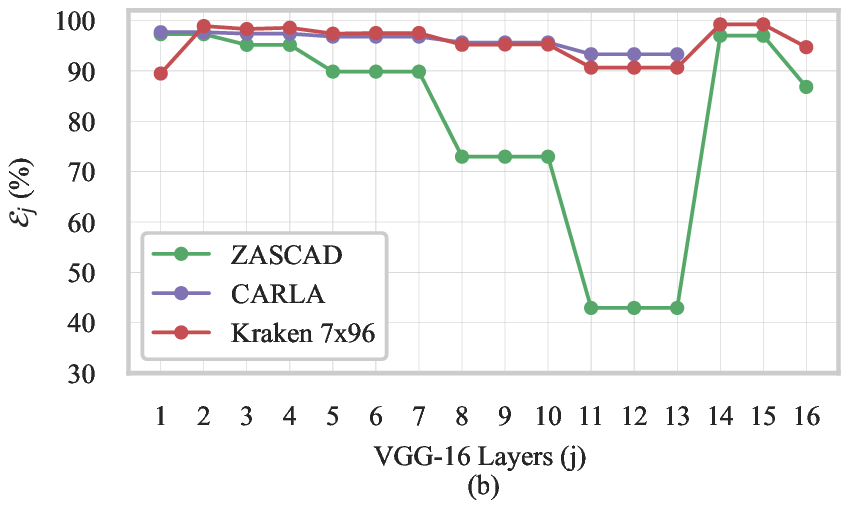}
    \includegraphics{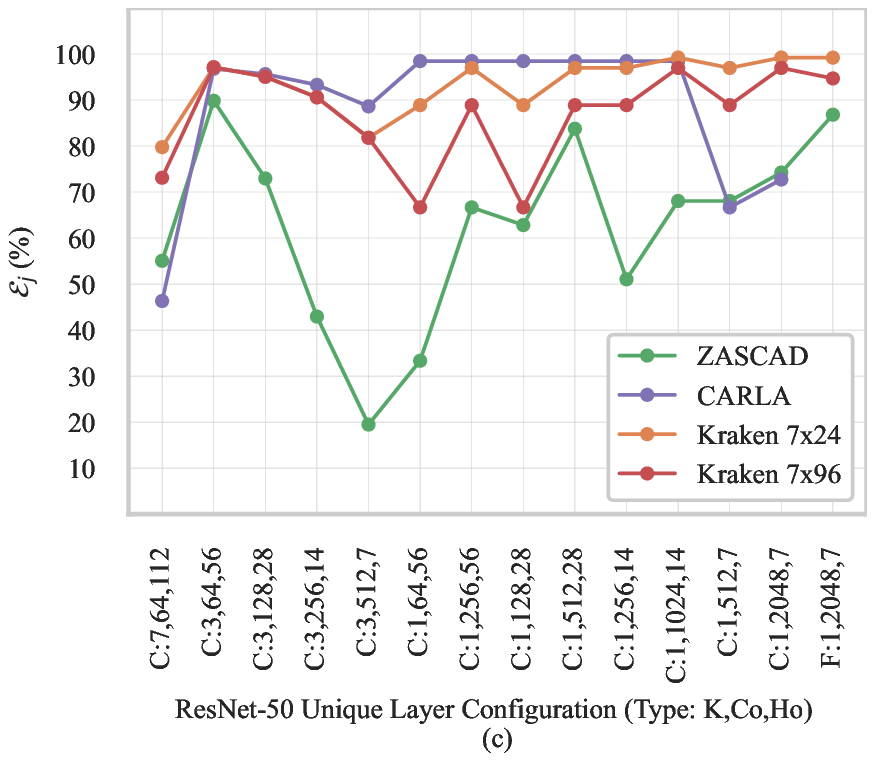}
    \includegraphics{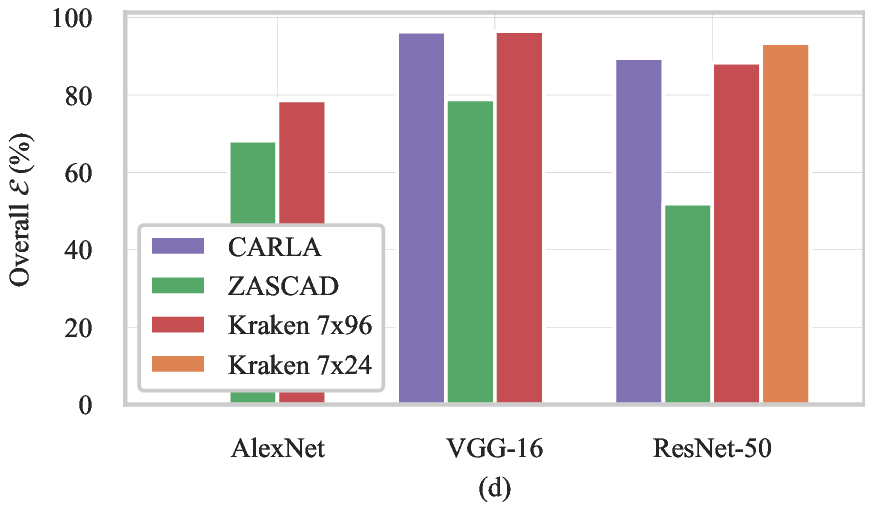}
  \end{center}
  \vspace*{-5mm}
  \caption{Comparison of performance efficiency: layer-wise ($\mathcal{E}_j$) on (a) AlexNet, (b) VGG-16, (c) ResNet-50, and (d) overall ($\mathcal{E}$) on the three CNNs.}
  \label{fig:eff}
\end{figure}

Figure \ref{fig:eff} presents the performance efficiencies of Kraken, calculated using the exact expressions derived in \eqref{eq:kraken_ej} and \eqref{eq:p_eff}, with that of the prior works calculated using the number of valid MACs (Table \ref{table:cnns}) and formulae presented in respective papers for the number of clock cycles, when processing each layer of AlexNet, VGG-16, and ResNet-50. Note that, for ResNet-50, only the layer configurations, where Kraken, CARLA, and ZASCAD demonstrate unique (non-repetitive) performance efficiency values, are presented.

We note that Eyeriss \cite{eyeriss} \cite{eyeriss_v2} and MMIE/ZASCAD \cite{zasca} have included the wasted operations associated with zero padding when presenting their performance (Gops) and energy efficiency whereas we follow CARLA \cite{carla} in ignoring such operations (see \eqref{eq:macs}). This is because, as Kraken and CARLA perform zero-padding without overhead, inclusion of zero-pads into the operation count yields unrealistic performance efficiencies above 100\%. Therefore, for a consistent comparison, these metrics of prior works have been recalculated using \#MAC$_\textbf{valid}$ (see \eqref{eq:macs} and Table \ref{table:cnns}), their throughputs (fps) and number of clock cycles. 

\subsubsection{Eyeriss (JSSC’17)} was introduced in \cite{eyeriss} as an array of $12 {\times} 14 = 168$ PEs. Each PE consists of a 224-word deep, 16-bit wide SRAM, a 41-word register bank, 4 FIFOs, 5 registers, 2 two-way multiplexers and a controller, in addition to the multiplier and the adder. This results in 60\% of the per-PE area and 47.9\% of the total area being utilized for PE scratchpads (SRAM and register bank), while only 9.4\% of the per-PE area being used for the multiplier and the adder. In Eyeriss v2 \cite{eyeriss_v2}, each PE is implemented using seven pipeline stages and five scratchpads, with 288 bytes of SRAM and 98.5 bytes of registers per PE, resulting in only 5.2\% of the per-PE area utilized for the two multipliers and adders. In contrast, Kraken's dataflow eliminates the need for scratchpads inside PEs, resulting in \vAreaPEMacPercent{}\% of the per-PE area is used by the multiplier and the accumulator, making it possible to pack \vNumPETimesEyeriss{}${\times}$ more PEs, and \vMemoryTimesEyeriss{}${\times}$ more memory (as global buffer) in \vAreaTimesEyeriss{}${\times}$ the area compared to Eyeriss, as shown in Table \ref{table:compare_conv}. We note that the area and power metrics of Eyeriss is presented from their fabricated chip while Kraken's metrics are post-synthesis. 

The PE array of Eyeriss is assigned work by a Network-on-Chip using either multicast or point-to-point data delivery dictated by their \emph{row-stationary} dataflow. Reconfiguration after each layer is done by serially feeding a 1794-bit scan chain, which takes about 100 $\mu s$. In addition to the relatively low utilization on processing clock cycles, the PE array is idle during reconfiguration and while data is being transferred from and to off-chip DRAM, resulting in low overall performance efficiencies of 63.6\% and 30.8\% for AlexNet and VGG-16. Meanwhile, Kraken takes just zero or one clock cycle (2.5 ns) to load the configuration data. In addition, reconfiguration and control paths are decentralized, and relatively smaller buffers are employed, eliminating the need to stall the engine during reconfiguration and data transfer. As a result, while Eyeriss achieves fewer memory accesses and higher arithmetic intensity, Kraken outperforms Eyeriss in terms of performance efficiency, throughput and latency as shown in Table \ref{table:compare_conv}.


\begin{figure}
  \centering
  \def\svgwidth{\columnwidth}
  \begin{center}
    \includegraphics{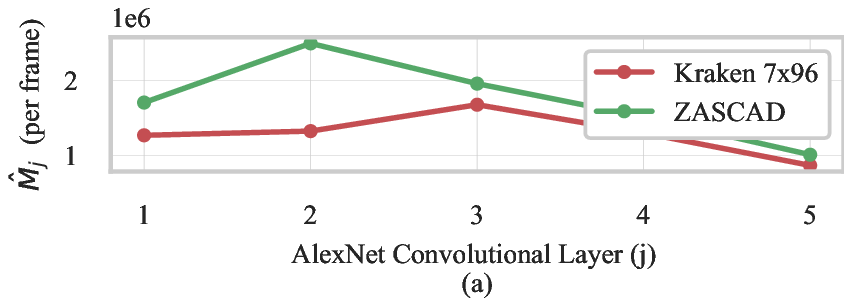}
    \includegraphics{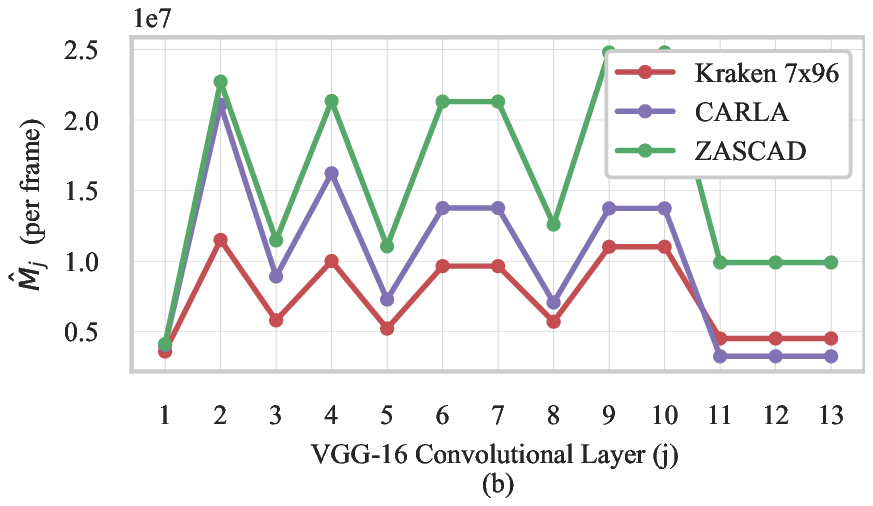}
    \includegraphics{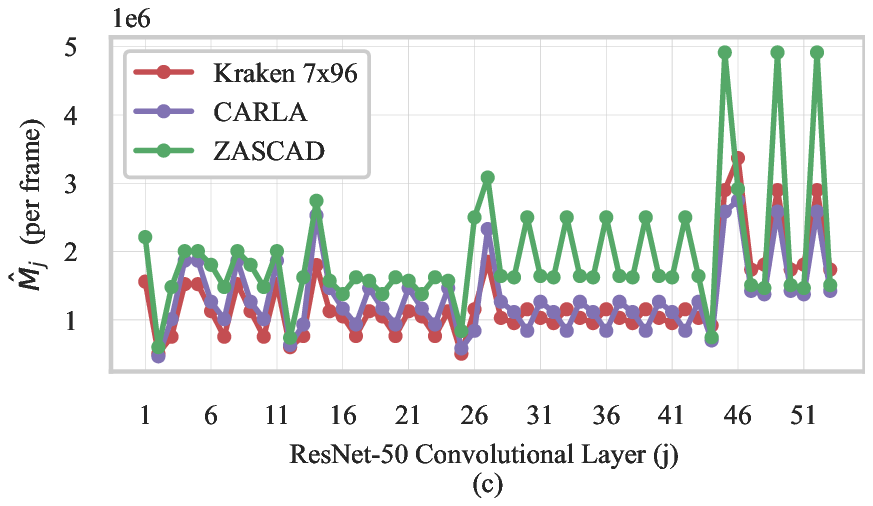}
    \includegraphics{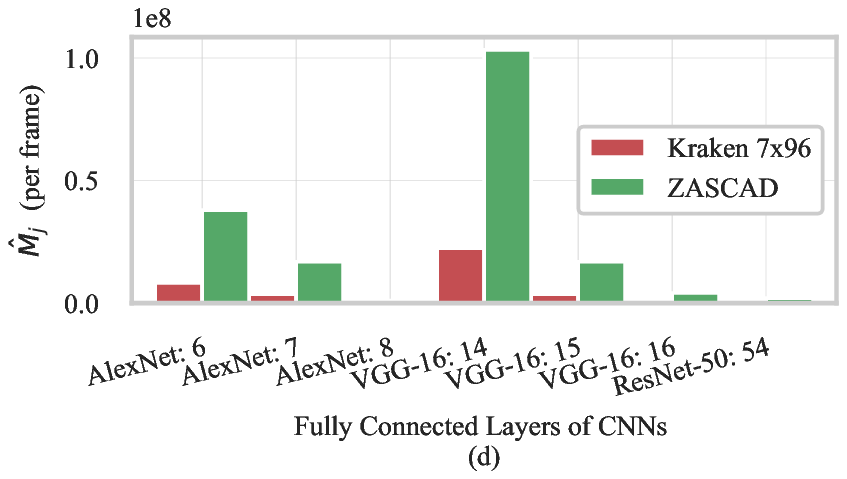}
    \includegraphics{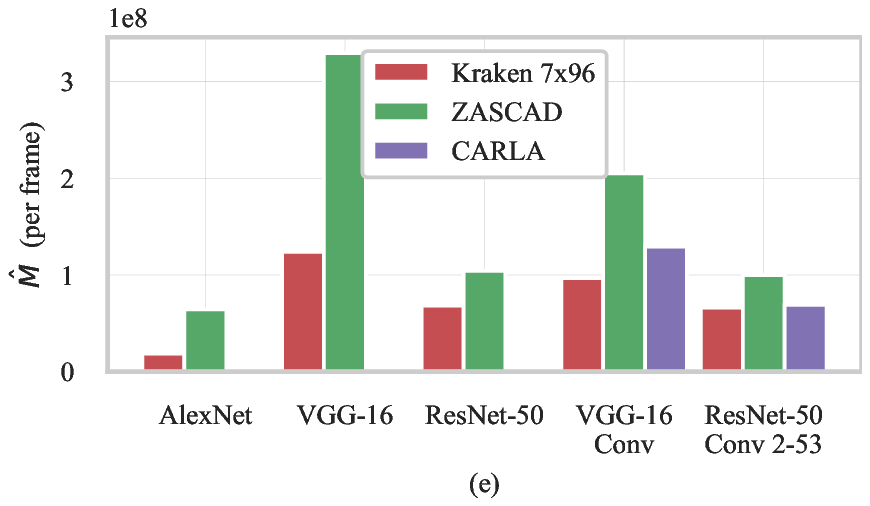}
  \end{center}
  \vspace*{-7mm}
  \caption{Comparison of number of memory accesses: on convolutional layers of (a) AlexNet, (b) VGG-16, (c) ResNet-50, (d) on fully-connected layers of each CNN, and (e) total on each CNN.}
  \label{fig:ma}
\end{figure}

\subsubsection{MMIE \texorpdfstring{\cite{multi_mode}}{citemmie} / ZASCAD \texorpdfstring{\cite{zasca}}{citezasca} (TCOMP'20)} is built as an array of 32 1-D reconfigurable tiles. The 6 PEs of each such tile dynamically group into effective tiles, each tasked with computing one output channel. This limits the reconfigurability to only a handful of $K,S$ combinations and leaves higher percentage of PEs idle on every clock for certain layers. Each of the 192 PEs of ZASCAD contains a 192 bytes of SRAM (64 words deep, 24 bits wide) to store the consecutive pixels to be used in vertical convolution ($\Sigma^{K_H}$). Each PE also has a corresponding 11-word register bank and an 11-way multiplexer in the tile's weight generator, most of which are unused for most $K,S$ combinations. This results in a much larger per-PE area. Kraken's novel dataflow performs vertical convolution via interleaved pixel shifting and simplifies PE design. Weights Rotator has two large SRAM banks, which memory compilers optimize better than hundreds of scattered smaller SRAMs, allowing Kraken to pack \vNumPETimesZascad{}${\times}$ more PEs and \vMemoryTimesZascad{}${\times}$ more memory in \vAreaTimesZascad{}${\times}$ the area.

We note that \cite{zasca} calculates the number of clock cycles needed for MMIE/ZASCAD to compute a layer, ignoring the extra iteration needed to compute $H_{out}{\times}W_{out}$ pixels in groups of $N$ ((11) in \cite{zasca}). Whereas this helps to report a much higher performance efficiency, it results in unrealistic estimates of fractional number of clock cycles for several layers of AlexNet. Since Kraken is evaluated considering all extra iterations, see \eqref{eq:qj}, \eqref{eq:t} and \eqref{eq:t}, for a fair comparison, Fig. \ref{fig:eff} demonstrates the performance efficiencies of MMIE/ZASCAD, using $\left \lceil{(H_{out}{\times}W_{out})/N}\right \rceil$ to accurately consider the under-utilized extra iterations as well. Furthermore, while MMIE reports high utilization factors (percentage of PEs active in a computational clock cycle), it wastes several clock cycles in a process called \emph{weights passing} when starting each new row, and is unable to perform computations when streaming out output pixels. This results in a low overall performance efficiency calculated using their reported clock cycles and valid MACs (see \eqref{eq:macs}). In contrast, Kraken fetches weights for the next iteration while rotating weights for the current one and streams out output pixels without stalling the engine. Therefore Kraken outperforms MMIE in performance efficiency (Fig. \ref{fig:eff}) for both convolutional and fully-connected layers of AlexNet, VGG-16, and ResNet-50.

While ZASCAD/MMIE accelerates fully-connected layers, it fails to reuse their weights, which dominate the energy consumption by being 94.3\% and 76.8\% of all memory accesses required to compute AlexNet and VGG-16, respectively (see Table \ref{table:cnns}). In contrast, Kraken processes $R$ batches in parallel using the loaded weights, when computing fully connected layers, resulting in a much fewer memory accesses per frame in both fully connected layers and overall network  as shown in Fig. \ref{fig:ma} and a much higher arithmetic intensity as presented in Table \ref{table:compare_fc}. Kraken significantly outperforms ZASCAD in every metric presented in Tables \ref{table:compare_conv} and \ref{table:compare_fc} due to more PEs operating at higher frequency, and a more efficient dataflow that maximizes data reuse and overall performance efficiency.

\subsubsection{CARLA (TCAS'21 \texorpdfstring{\cite{carla}}{citecarla})} is built as an array of 65 cascaded convolutional units (CUs), where the first 64 contain 3 PEs and the last CU contains 4 PEs. Four distinct dataflows are employed to achieve performance efficiency in those layers, requiring each of the 196 PEs to have a pair of 224 word SRAMs and an input register, and each CU to have a 4-way mux, two 3-way muxes, two 2-way muxes and two registers. Fully-connected layers are not processed, and their performance is not measured. In contrast, Kraken employs a single, uniform dataflow which is able to optimally process any convolutional layer, fully-connected layer or matrix product to outperform CARLA in overall metrics, eliminating the need for complex PEs and data routing in its architecture. Consequently, Kraken's implementation packs \vNumPETimesCarla{}${\times}$ PEs and \vMemoryTimesCarla{}${\times}$ SRAM for much better data reuse, in just \vAreaTimesCarla{}${\times}$ the area, running at \vFreqTimesCarla{}${\times}$ the frequency, resulting in a peak performance of \vGopsPerAreaTimesCarla{}${\times}$ more Gops/\si{mm^2} and \vGopsPerWTimesCarla{}${\times}$ more Gops/W compared to CARLA, as presented in Tables \ref{table:compare_conv} and \ref{table:compare_fc}.

The architecture of CARLA has been tailored for the convolutional layers of VGG and ResNet CNNs, such that SRAM depth, number of PEs and number of CUs are factors of the dimensions of those networks. Fig. \ref{fig:eff}  demonstrates the Kraken $7{\times}24$, similarly optimized for only these CNNs, outperforming CARLA with 93.3\% performance efficiency in the convolutional layers of ResNet-50 compared to CARLA's \vResCarlaConvPEf{}\%. However, the $R,C{=}\vR{},\vC{}$ configuration is implemented for being efficient over all kinds of CNNs (including AlexNet), while requiring fewer memory accesses.

CARLA's PE utilization factor (PUF: percentage of PEs active in a computational clock cycle) of 98.46\% reported for $3{\times}3$ convolutional layers is from a reported formula that has been simplified with certain assumptions, which do not hold for all considered $3{\times}3$ layers. When using the accurate formula presented in \cite{carla}, some degradation is observed in PUF. While the tailored architecture allows CARLA to achieve over 90\% utilization in $3{\times}3$ and the initial $1{\times}1$ layers of ResNet-50, its performance efficiency drops to 45\% for $7{\times}7$ and 73\% for the latter $1{\times}1$ layers. 

Due to poor utilization over layers with large filter sizes, CARLA is not evaluated on AlexNet, whose $11{\times}11$ and $5{\times}5$ convolutional layers contain 49\% of its computations. In contrast, Kraken processes convolutional layers with larger filter sizes with acceptable performance efficiencies, due to its elastic grouping. As a result, Kraken $7{\times}24$ and $\vR{}{\times}\vC{}$ achieve performance efficiencies of \vResFirstAltPEf{}\% and \vResFirstPEf{}\% compared to CARLA's 45\% on the first convolutional layer of ResNet-50. 

As demonstrated in in Figs. \ref{fig:eff} and \ref{fig:ma}, and Table \ref{table:compare_conv}, the uniform dataflow and generalized architecture of Kraken $\vR{}{\times}\vC{}$ outperforms the multiple dataflows and the tailored architecture of CARLA in overall performance efficiency (except for ResNet-50), arithmetic intensity, the number of memory accesses performance, and energy efficiency, while using simpler and smaller PEs resulting in a much better Gops/area and Gops/W performance.
\section{Acknowledgment}

We thank Rukshan Wickramasinghe for his assistance in developing an early version of the engine, and Prof. Rohan Munasinghe (UoM), Dr. Thayaparan Subramaniam (UoM), and Udara De Silva (FIU) for helpful discussions. We also thank Arm Ltd. for providing the PDKs and memory compilers through the Arm Academic Access program.

\section{Conclusion}

This paper presents the first generation work to introduce the Kraken architecture and its corresponding dataflow for the inference of dense DNNs, which maximally exploits data reuse in weights, inputs, and outputs with a bare-bones PE design. Furthermore, Kraken's architecture features dynamic, distributed reconfiguration that takes at most one clock cycle and propagates with data, elastically grouping its cores on the fly, resulting in high overall performance efficiency. A detailed performance analysis is presented, deriving key metrics as exact functions of static parameters, which are then optimized to obtain the static configuration that is then implemented in TSMC 65-nm GP CMOS technology. Kraken's uniform dataflow is shown to be able to process convolutional layers, fully-connected layers, and matrix products of any shape, outperforming the state-of-the-art in overall performance efficiency, number of memory accesses, and arithmetic intensity, with \vGopsPerAreaTimesCarla{}${\times}$ more Gops/mm$^2$ and \vGopsPerWTimesCarla{}${\times}$ more Gops/W. The implemented system at \vConvFreqMhz{} MHz is shown to have a performance of up to \vGopsPeak{} Gops, processing the convolutional layers of AlexNet, VGG-16, and ResNet-50 at a throughput of \vAlexConvFps{}, \vResConvFps{}, and \vVggConvFps{} frames/s, respectively.

\bibliography{references,IEEEabrv}
\bibliographystyle{ieeetr}

\begin{IEEEbiography}[{\includegraphics[width=1in,clip,keepaspectratio]{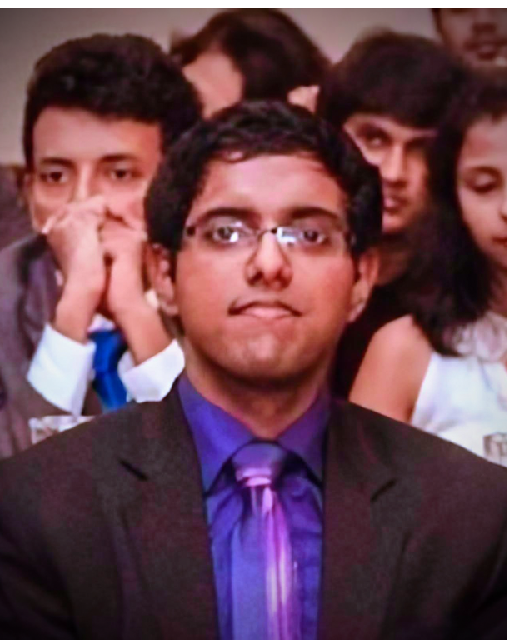}}]{G Abarajithan} was born in Peradeniya, Sri Lanka in 1995. He received the B.S degree in electronics and telecommunications engineering from the University of Moratuwa (UoM), Moratuwa, Sri Lanka, in 2020 with first-class honors. He is currently serving as an RTL Design Engineer at Lemurian Labs (Canada), and a consultant at UoM. His research interests include computer architecture, system-on-chip design, and hardware acceleration.
\end{IEEEbiography}

\begin{IEEEbiography}[{\includegraphics[width=1in,clip,keepaspectratio]{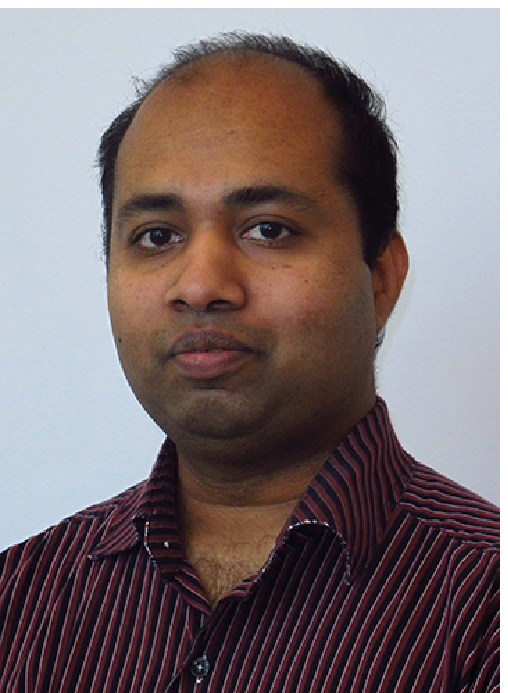}}]{Chamira U. S. Edussooriya}
received the B.Sc.Eng. degree in Electronic and Telecommunication Engineering (first class honors) from the University of Moratuwa, Moratuwa, Sri Lanka, in 2008, and the M.A.Sc. and the Ph.D. degrees in Electrical Engineering from the University of Victoria, Victoria, BC, Canada, in 2012 and 2015, respectively. He has been a Senior Lecturer at the Department of Electronic and Telecommunication Engineering, University of Moratuwa since January 2016, and a Courtesy Post-Doctoral Associate at the Department of Electrical and Computer Engineering, Florida International University, Miami, FL, USA since December 2019. He briefly visited the Incheon National University, Incheon, South Korea in March 2019 and the Florida International University in March and April 2019.

He has been an executive committee member of the IEEE Sri Lanka Section and the Chair of the Educational Activities Committee in 2021. Furthermore, he is the founding chair of the IEEE Sri Lanka Section Signal Processing Society Chapter and the founding faculty advisor of the IEEE Signal Processing Society Student Branch Chapter at the University of Moratuwa. He served as a publication co-chair and the special sessions chair of Moratuwa Engineering Research Conference in 2020 and 2021, respectively. His current research interests include analysis and design of low-complexity multidimensional digital filters, 4-D light field and 5-D light field video processing, array signal processing and machine learning techniques for multi-dimensional signal processing. His research contributions include co-invention of the class of multi-dimensional filters (called 5-D depth-velocity filters) applied to light field videos.    

\end{IEEEbiography}


\end{document}